\documentclass[fleqn,usenatbib]{mnras}

\usepackage[T1]{fontenc}
\usepackage{ae,aecompl}

\usepackage{graphicx}   
\usepackage{amsmath}    

\usepackage{amssymb}    

\usepackage{multicol}

\usepackage{pdflscape}

\usepackage{eso-pic}

\AddToShipoutPictureBG*{%
  \AtPageUpperLeft{%
    \hspace{0.75\paperwidth}%
    \raisebox{-3.5\baselineskip}{%
      \makebox[0pt][l]{\textnormal{DES 2017-0293}}
}}}%

\AddToShipoutPictureBG*{%
  \AtPageUpperLeft{%
    \hspace{0.75\paperwidth}%
    \raisebox{-4.5\baselineskip}{%
      \makebox[0pt][l]{\textnormal{FERMILAB-PUB-19-104-AE}}
}}}%

\title[Brown dwarf census with DES Y3]{Brown dwarf census with the Dark Energy Survey year 3 data and the thin disk scale height of early L types}

\author[DES Collaboration]{
\parbox{\textwidth}{
\Large
A.~Carnero~Rosell,$^{1,2,3}$
B.~Santiago,$^{4,2}$
M.~dal Ponte,$^{4}$
B.~Burningham,$^{5}$
L.~N.~da Costa,$^{2,3}$
D.~J.~James,$^{6,7}$
J.~L.~Marshall,$^{8}$
R.~G.~McMahon,$^{9,10}$
K.~Bechtol,$^{11,12}$
L.~De Paris,$^{4}$
T.~Li,$^{13,23}$
A.~Pieres,$^{2,3}$
T.~M.~C.~Abbott,$^{14}$
J.~Annis,$^{13}$
S.~Avila,$^{15}$
G.~M.~Bernstein,$^{16}$
D.~Brooks,$^{17}$
D.~L.~Burke,$^{18,19}$
M.~Carrasco~Kind,$^{20,21}$
J.~Carretero,$^{22}$
J.~De~Vicente,$^{1}$
A.~Drlica-Wagner,$^{14,23}$
P.~Fosalba,$^{24,25}$
J.~Frieman,$^{13,23}$
J.~Garc\'ia-Bellido,$^{15}$
E.~Gaztanaga,$^{24,25}$
R.~A.~Gruendl,$^{20,21}$
J.~Gschwend,$^{2,3}$
G.~Gutierrez,$^{13}$
D.~L.~Hollowood,$^{26}$
M.~A.~G.~Maia,$^{2,3}$
F.~Menanteau,$^{20,21}$
R.~Miquel,$^{27,22}$
A.~A.~Plazas,$^{28}$
A.~Roodman,$^{18,19}$
E.~Sanchez,$^{1}$
V.~Scarpine,$^{13}$
R.~Schindler,$^{19}$
S.~Serrano,$^{24,25}$
I.~Sevilla-Noarbe,$^{1}$
M.~Smith,$^{29}$
F.~Sobreira,$^{30,2}$
E.~Suchyta,$^{31}$
M.~E.~C.~Swanson,$^{21}$
G.~Tarle,$^{32}$
V.~Vikram,$^{33}$
and A.~R.~Walker$^{14}$
\begin{center} (DES Collaboration) \end{center}
}
\\
\\
Affiliations are listed at the end of the paper.
}

\date{Accepted XXX. Received YYY; in original form ZZZ}

\pubyear{2019}
\begin{document}
\label{firstpage}

\maketitle


\begin{abstract}
In this paper we present a catalogue of 11,745 brown dwarfs with spectral types ranging from L0 to T9, photometrically classified using data from the Dark Energy Survey (DES) year 3 release matched to the Vista Hemisphere Survey (VHS) DR3 and Wide-field Infrared Survey Explorer (WISE) data, covering $\approx 2,400 \ deg^{2}$ up to $i_{AB}=22$. The classification method follows the same photo-type method previously applied to SDSS-UKIDSS-WISE data. The most significant difference comes from the use of DES data instead of SDSS, which allow us to classify almost an order of magnitude more brown dwarfs than any previous search and reaching distances beyond 400 parsecs for the earliest types. Next, we also present and validate the \textit{GalmodBD} simulation, which produces brown dwarf number counts as a function of structural parameters with realistic photometric properties of a given survey. We use this simulation to estimate the completeness and purity of our photometric LT catalogue down to $i_{AB}=22$, as well as to compare to the observed number of LT types. We put constraints on the thin disk scale height for the early L population to be around 450 parsecs, in agreement with previous findings. For completeness, we also publish in a separate table a catalogue of 20,863 M dwarfs that passed our colour cut with spectral types greater than M6. Both the LT and the late M catalogues are found at \url{https://des.ncsa.illinois.edu/releases/other/y3-mlt}. 
\end{abstract}

\begin{keywords}
Catalogues, Surveys, brown dwarfs, infrared: stars, techniques: photometric
\end{keywords}

\section{Introduction}

Ultra-cool dwarfs are mostly sub-stellar objects (brown dwarfs, BDs) with very cool ($T_{\rm eff} < 2,300 \ K$) atmospheres with spectral types later than M7, including the L, T and Y sequences.  Their spectra are typified by the effects of clouds and deep molecular absorption bands. In L dwarfs ($2,200 \gtrsim T_{\rm eff} \gtrsim 1,400 \ K$) clouds block radiation from emerging from deep in the atmosphere in the opacity windows between molecular absorption bands, narrowing the pressure range of the observed photosphere, and redistributing flux to longer wavelengths giving these objects red near-infrared colours. The transition to the T sequence (LT transition, $T_{\rm eff} \sim 1,200-1,400 \ K$) is driven by the disappearance of clouds from the near-infrared photosphere, leading to relatively bluer colours. This is accompanied by the transition from CO (L dwarfs) to CH$_{4}$ (T dwarfs) dominated carbon chemistry.  At cooler temperatures ($T_{\rm eff} < 500 \ K$), the development of the Y dwarf sequence is thought to be driven by the emergence of clouds due to sulphide and chloride condensates, as well as water ice \citep[e.g.][]{leggett2013,leggett2015,skemer2016}.   

BDs never achieve sufficient core temperatures to maintain main-sequence hydrogen fusion. Instead, they evolve from relatively warm temperatures at young ages to ever cooler temperatures with increasing age as they radiate the heat generated by their formation. As a result, the late-M and early-L~dwarf regime includes both young, high-mass, brown dwarfs and the lowest mass stars. The latter can take several hundred million years to reach the main sequence. Objects with late-L, T and Y spectral types are exclusively substellar. In this work, we focus on L and T dwarfs, and for brevity refer to this group as BDs.  

BDs have very low luminosity, especially the older or lower mass ones. Their mass function, star formation history (SFH) and spatial distribution are still poorly constrained, and the evolutionary models still lack details, especially the lowest-masses and old ages. They are supported at their cores by degenerate electron pressure and, because the degeneracy determines the core density (instead of the Coulomb repulsion), more massive BDs have smaller radii. BDs are actually partially-degenerate, in the sense that while in their atmospheres reign thermal pressure, somewhere in their interior there must be a transition from degenerate electron to thermal pressure. 

The current census covers an age range from a few million years \citep{liu2013,gagne2017} to $>10 \ Gyr$ halo members \citep{bur03,zhang2017}, and spans the entire mass interval between planetary and stellar masses. The diverse range of properties that these objects display reflects the continual luminosity and temperature evolution of these partially-degenerate objects. As a numerous and very long-lived component of our Galaxy, these continually evolving objects could be used to infer structural components of the Milky Way, and tracing the low-mass extreme of star formation over cosmic timescales. However, studies of L dwarfs have typically been restricted to the nearest $100 \ pc$, while T dwarfs are only known to distances of a few tens of parsecs.

\begin{table*}
\centering
\caption{Summary of the major photometric brown dwarf census so far. In the third column we show the colour selection in its original systems, according to the surveys used (indicated in the first column). For comparison, we also show the selection used in this paper, where we classify more than $10,000$ BDs up to $z_{AB}=22$.}
\label{tab:prevsurveys}

\begin{tabular}{| c | c | c | c | c | }
    
    \hline \hline
    
Reference &  Area &  Selection & Spectral Types & Number   \\ 
 (Surveys) & ($deg^{2}$) &&& \\
\hline \hline  
                              
~\citet{chiu2006}   & 3,526 & $(i-z)_{AB} > 2.2$, & LT & 73 \\ 

(SDSS) &&$z_{AB}<20.4$&& \\ \hline

~\citet{sch10}   & 11,000  & $(i-z)_{AB}>1.4$  & L & 484 \\ 

(SDSS, 2MASS) &&&& \\ \hline

~\citet{albert2011}   & 780 &  $(i-z)_{AB} > 2.5$ & T & 37 \\ 

(CFHT) &&&& \\ \hline

~\citet{kirk2011}   & All sky &  T5: $(W1-W2)_{Vega}>1.5$,  $(W2-W3)_{Vega}<3$.   & LT & 103 \\ 

(WISE) &&L: $(W1-W2)_{Vega}>0.$&& \\ \hline

~\citet{lodieu2012} &  675 &  $(W1-W2)_{Vega}>1.4$,  $(J-W2)_{Vega} >1.9$  & T & 13 \\ 

(VHS, WISE) &&&& \\ \hline

~\citet{DJ13}  &  500 &  $J_{Vega}<18.1$  & LT & 63 \\ 

(UKIDSS) &&&& \\ \hline

~\citet{Bur13}  & 2,000  & $(z_{AB}-J_{Vega}) > 2.5,$   & T & 76 \\ 

(UKIDSS) &&$J_{Vega}<18.8$&& \\ \hline
 
~\citet{skr2016}  & 3,070 &  $(Y-J)_{Vega} > 0.8,$   & LT & 1,361 \\ 

(SDSS, UKIDSS, WISE) &&$J_{Vega}<17.5$&& \\ \hline

~\citet{Sor18}  & 130 &  $1 < (i-z)_{AB} < 2.0$, $0.75< (z-Y)_{AB} < 1.0,$   & L & 3,665 \\ 

(HSC) &&$z_{AB}<24$&& \\ \hline

This paper (2019) & 2,400 & $(i-z)_{AB} >1.2$, $(z-Y)_{AB} > 0.15$, $(Y_{AB}-J_{Vega}) > 1.6$,  & LT & 11,745 \\ 

(DES, VHS, WISE) &&$z_{AB}<22$&& \\ \hline

    \end{tabular}
   \end{table*}                                

The era of digital wide-field imaging surveys has allowed the study of brown dwarfs to blossom, with thousands now known in the solar neighbourhood. But this collection is heterogeneous and very shallow and therefore, not suitable for large-scale statistical analysis of their properties. The new generation of deep and wide surveys (DES \citep{des16}, VHS \citep{mcmahon13}, UKIDDS \citep{lawrence07}, LSST \citep{abell2009lsst}, HSC \citep{Miy18}) offer the opportunity to place the brown dwarf population in their Galactic context, echoing the transition that occurred for M dwarfs with the advent of the Sloan Digital Sky Survey \citep[SDSS;][]{boch07,west11}. These surveys should be able to create homogeneous samples of BDs to sufficient distance to be suitable for various applications, such as kinematics studies \citep{fah09,fah12,sch10,smi14}, the frequency of binary systems in the LT population \citep{bur06c,bur07,luh12}, benchmark systems \citep{pin06,Bur13,mar17}, the search for rare or unusual objects \citep{bur03,fol07,loo08,skr2016}, and the study of Galactic parameters \citep{rya05,jur08,Sor18}. In this latter case, we will also need simulations to confront observed samples. Realistic simulations will benefit from improved spectral type-luminosity and spectral type-local density relations in the solar neighbourhood.

The UKIRT Infrared Deep Sky Survey \citep[UKIDSS; ][]{lawrence07} imaged $4000 \ deg^{2}$ in the $Y,J,H,K$ filter passbands, and provided discovery images for over 200 T dwarfs, making it one of the principal contributors to the current sample of LT dwarfs, particularly at fainter magnitudes \citep[e.g.][]{bur10,Bur13,bur18}. Experience gained through the exploitation of UKIDSS has demonstrated that significant amounts of 8m-class telescope time are required to spectroscopically classify samples of 10s to 100s of LT dwarfs within $100 \ pc$. For example, total observation times of 40 - 60 minutes were needed to obtain low-resolution spectra of $J \sim 18.5$ targets at a signal-to-noise ratio (SNR) = 20, necessary for spectral classification \citep{Bur13}. Obtaining homogeneous samples to the full depth available in the new generation of surveys is thus only feasible today through photometric classification. Such an approach was demonstrated in~\citet{skr14} and ~\citet{skr2016}, where they obtained a sample of more than 1000 LT dwarfs, independent of spectroscopic follow-up, using $i,z,Y,J,H,K_{s},W1,W2$ from $SDSS \cap UKIDDS \cap WISE$.

A summary of surveys that attempted to select BDs candidates photometrically can be found in Table~\ref{tab:prevsurveys}. We identify two approaches: one based on a colour selection in optical bands and another with colour selection in the near-infrared. In the first case, a common practice is to apply a cut on $(i-z)$. For example, \citet{sch10} apply a cut at $(i-z)_{SDSS}>1.4$, while \citet{chiu2006} cut at $(i-z)_{SDSS} > 2.2$ to select T types. This latter cut would be interesting to study the transition between L and T types, but not for a complete sample of L types. In any case, since the $i,z$ bands in DES are not precisely the same as in SDSS, we expect changes in our nominal colour cuts. 

When infrared bands are available, it is common to make the selection on $J$ band. For example, \citet{skr2016} apply a cut on Vega magnitudes $(Y-J)_{UKIDSS,vega} > 0.8$, which, in our case, would translate into $(Y_{AB}-J_{vega})_{UKIDSS} > 1.4$. \citet{Bur13} search for T types, applying a cut at $z_{AB}-J_{Vega} > 2.5$. Again, the UKIDSS filters are not exactly the same as DES or VHS, so we expect these cuts to change when applied to our data. 

In this paper, we follow the photo-type methodology of \citet{skr14} to find and classify L and T dwarfs in the $DES \cap VHS \cap AllWISE$ system (The AllWISE program was built by combining data from the WISE cryogenic and NEOWISE post cryogenic phases). We can go to greater distances due to the increased depth in the DES optical bands $i,z$ in comparison with SDSS while maintaining high completeness in the infrared bands, needed for a precise photometric classification. In fact, the optical bands can drive the selection of L dwarfs, as demonstrated here, improving upon previous photometric BDs searches. In the case of T dwarfs, the infrared bands are the limiting ones, and therefore, our sample will have a similar efficiency in that spectral regime in comparison with previous surveys. 

The methodology is based on three steps: first a photometric selection in colour space $(i-z)$, $(z-Y)$, $(Y-J)$ is done; second, a spectral classification is performed by comparing observed colours in $i,z,Y,J,H,K_{s},W1,W2$ to a set of colour templates for various spectral types, ranging from M1 to T9. These templates are calibrated using a sample of spectroscopically confirmed ultra-cool dwarfs (MLT). Finally, we remove possible extragalactic contamination with the use of a galaxy template fitting code, in particular, we use \verb+Lephare+ photo-z code\footnote{\url{http://www.cfht.hawaii.edu/~arnouts/LEPHARE/lephare.html}} \citep{Arn99, Ilb06}. 

After completion of a homogeneous sample of LT dwarfs, we proceed to measure the thin disk scale height ($h_{z,thin}$). Unfortunately, current simulations present many inconsistencies with observations and are not trustworthy. Therefore, we also introduce a new simulation which computes expected number counts of LT dwarfs and creates synthetic samples following the properties of a given survey. We have called it \textit{GalmodBD}. Finally, we can compare the output of the simulation for different formation scenarios, to the number of BDs found in the sample footprint, placing constraints on $h_{z,thin}$ and other fundamental parameters.

 
In Section~\ref{sec:data} we describe the data used in this paper ($DES \cap VHS \cap AllWISE$) and how these samples were matched and used through the analysis. In Section~\ref{sec:calib} we detail the samples we have used to define our colour selection as well as to create the colour templates that will feed the classification and the \textit{GalmodBD} simulation.

In Section~\ref{sec:targetselection} we show the colour based target selection scheme we have defined and how it compares with the previous analysis, in particular to ~\citet{skr14}. In Section~\ref{sec:classif_y3} we explain our classification methodology and how we apply it to the DES data in Section~\ref{mainclassif}. In Section~\ref{app:catexplanation} we detail the public catalogue. In Sections~\ref{sec:simulation} and ~\ref{sec:synth} we introduce the \textit{GalmodBD} simulation and how we tune it to our data in Section~\ref{sec:galmodbdindes}. In Section~\ref{sec:galmobdresults} we present results of running the simulation. In Section~\ref{sec:systematicsgalmod} we use the \textit{GalmodBD} to study the completeness of our photometric selection and other systematics applying to the number of BDs detected. Finally, in Section~\ref{sec:sim_des} we compare results from the \textit{GalmodBD} simulation to our data, placing constraints on the thin disk scale height of L types. 

Through the paper, we will use the photometric bands $i,z,Y$ from DES in AB magnitudes, $J,H,K_{s}$ from VHS in Vega and $W1,W2$ from AllWISE in Vega. The use of Vega or AB normalisation is not important since it is only a constant factor when calculating colours. 

\section{the data}
\label{sec:data}

In this section we present the three photometric data sets used in this analysis: the Dark Energy Survey (DES), with 5 filters covering from 0.38 to 1 $\mu m $ (from which we use $i,z,Y$), VHS, with 3 filters covering from 1.2 to 2.2 $\mu m$ and AllWISE $W1$ and $W2$, at 3.4 and 4.6 $\mu m$. Finally, in Subsection~\ref{subsec:desvhswise} we detail the catalogue matching process and quality cuts.

\subsection{The Dark Energy Survey (DES)}
\label{subsec:des}

DES is a wide-field optical survey in the $g,r,i,z,Y$ bands, covering from $3800 \ \text{\AA}$ to $\sim 1 \ \mu m$. The footprint was designed to avoid extinction contamination from the Milky Way as much as possible, therefore pointing mostly towards intermediate and high Galactic latitudes. The observational campaign ended on the 9th of January 2019. The final DES data compromises 758 nights of observations over six years (from 2013 to 2019).

In this paper we use DES year 3 (Y3) data, an augmented version of the Data Release 1 \citep[The DR1;][]{abb18}\footnote{\url{https://des.ncsa.illinois.edu/releases/dr1}} which contains all the observations from 2013 to 2016. DES DR1 covers the nominal 5,000 $deg^2$ survey region. The median coadded catalogue depth for a $1.95\arcsec$ diameter aperture at $S/N = 10$ is $i \sim 23.44$, $z \sim 22.69$, and $Y \sim 21.44$. DR1 catalogue is based on coaddition of multiple single epochs \citep{mor18} using \verb+SExtractor+ \citep{Ber99}. 
The DES Data Management (DESDM) system also computes morphological and photometric quantities using the ``Single Object Fitting" pipeline (SOF), based on the \verb+ngmix+\footnote{\url{https://github.com/esheldon/ngmix}} code, which is a simplified version of the MOF pipeline as described in \citep{drl18}. The SOF catalogue was only calculated in $g,r,i,z$, but not in $Y$, therefore we use \verb+SExtractor+ $Y$ measurement from DR1 data. All magnitudes have been corrected by new zeropoint values produced by the collaboration, improving over those currently published in the DR1 release.

Information about the mean wavelength of each passband and magnitude limit at $5\sigma$ (defined as the mode of the magnitude distribution, as cited above, for a catalogue with $S/N>5\sigma$) is given in Table~\ref{tab:errcurve}. In Fig.~\ref{fig:y3footprint} we show the DES footprint with coverage in $i,z,Y$. It has an area = $5019 \ deg^{2}$ and all coloured areas represent the DES footprint.

To ensure high completeness in the $i$ band and infrared bands with sufficient quality, we impose a magnitude limit cut of $z<22$ with a detection of $5\sigma$ at least in the $z$ and $Y$ magnitudes. To avoid corrupted values due to image artefacts or reduction problems, we also apply the following cuts: \verb+SEXTRACTOR_FLAGS_z,Y = 0+: ensures no reduction problems in the $z$ and $Y$ bands. \verb+IMAFLAGS_ISO_i,z,Y = 0+: ensures the object has not been affected by spurious events in the images in $i,z,Y$ bands.

\subsection{VISTA Hemisphere survey (VHS)}
\label{subsec:vhs}

The  VISTA  Hemisphere  Survey \citep[VHS,][]{mcmahon13} is an infrared photometric survey aiming to observe 18,000 $deg^2$ in the southern hemisphere, with full overlap with DES in two wavebands, $J$ and $K_{s}$, to a depth $J_{AB} \sim 21.2$, $K_{s,AB} \sim 20.4$ at 5$\sigma$ for point sources ($J_{Vega} \sim 20.3, \ K_{s,Vega} \sim 18.6$, respectively) and partial coverage in $H$ band, with depth $H_{AB} \sim 19.85$ at $5\sigma$ ($H_{Vega} \sim 18.5$). The VHS uses the 4m VISTA telescope at ESO Cerro Paranal Observatory in Chile with the VIRCAM camera \citep{dalton06}. The data were downloaded by the DESDM system from the ESO Science Archive Facility \citep{cross12} and the VISTA Science Archive\footnote{\url{http://horus.roe.ac.uk/vsa/coverage-maps.html}}. 

The VHS DR3 covers 8,000 $deg^2$ in $J,K_{s}$ of observations from the year 2009 to 2013, from which a smaller region overlaps with DES, as shown in brown in Fig.~\ref{fig:y3footprint}. The coverage area in common with DES is $2374 \ deg^{2}$ for the $J,K_{s}$ filters, whereas addition of the $H$ band reduces this to $1331 \ deg^{2}$ (shown as light brown in Fig.~\ref{fig:y3footprint}).

In this paper, we use only sources defined as primary in the VHS data. We also impose 5$\sigma$ detection in $J$; whenever $H,K_{s}$ are available, we use them for the spectral classification (see Section \ref{sec:classif_y3}). 

We use \verb+apermag3+ as the standard VHS magnitude, in the Vega system, defined as the magnitude for a fixed aperture of $2\arcsec$. In Table~\ref{tab:errcurve} we show the summary of the filters and magnitude limits for VHS. VHS magnitude limits are in AB, even though we work in Vega throughout the paper. We use the transformation given by the VHS collaboration \footnote{\url{http://casu.ast.cam.ac.uk/surveys-projects/vista/technical/filter-set}}: $J_{AB}=J_{Vega} + 0.916, \ H_{AB} = H_{Vega} + 1.366, \ K_{s,AB}=K_{s,Vega}+1.827$.

\subsection{AllWISE}
\label{subsec:wise}

We also use AllWISE\footnote{\url{http://wise2.ipac.caltech.edu/docs/release/allwise/}} data, a full sky infrared survey in 3.4, 4.6, 12, and 22 $\mu m$, corresponding to $W1,W2,W3,W4$, respectively. AllWISE data products are generated using the imaging data collected and processed as part of the original WISE \citep{wri10} and NEOWISE \citep{mai11} programs. 

Because LT colours tend to saturate for longer wavelengths, we will make use only of $W1$ and $W2$. The AllWISE catalogue is $>$95\% complete for sources with $W1 < 17.1$ and $W2 < 15.7$ (Vega). 

In Table~\ref{tab:errcurve} we also show the properties of the AllWISE filters and magnitude limits. Magnitudes are given in AB using the transformations given by the collaboration \footnote{\url{http://wise2.ipac.caltech.edu/docs/release/allsky/expsup/sec4_4h.html}}: $W1_{AB} = W1_{Vega} + 2.699, \ W2_{AB}=W2_{Vega} + 3.339$.

Since our primary LT selection criteria do not use $W1$ and $W2$ magnitudes, we do not demand the availability of magnitudes in these bands when selecting our candidate sample. In other words, if a source has no data from AllWISE, we still keep it and flag their $W1$ and $W2$ magnitudes as unavailable in the classification.

\subsection{Combining DES, VHS and AllWISE data}
\label{subsec:desvhswise}

We first match DES to VHS with a matching radius of $2\arcsec$, and with the resulting catalogue, we repeat the same process to match to the AllWISE catalogue, using DES coordinates. The astrometric offset between DES and VHS sources was estimated in \citet{bane15}, giving a standard deviation of $0.18\arcsec$. For sources with significant proper motions, this matching radius may be too small. For instance, an object at $10 \ pc$ distance, moving at $30 \ km/s$ in tangential velocity, has a proper motion of $0.6  \arcsec/yr$. So, a matching radius of $2\arcsec$ will work except for the very nearby $(<= 6pc)$ or high-velocity $(> 50 km/s)$ cases, given a 2-year baseline difference in the astrometry. In fact, high velocity nearby BDs are interesting, since they may be halo BDs going through the solar neighbourhood. A small percentage of BDs will be missing from our catalogue due to this effect. In Section~\ref{sec:systematicsgalmod} we quantify this effect.

\begin{figure}
 \includegraphics[width=\columnwidth]{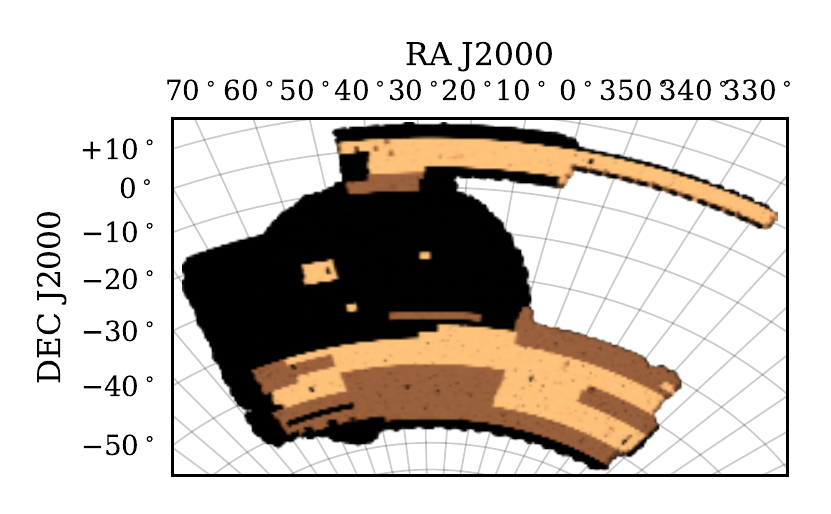}
 \caption{The footprint of DES Y3 data considering the intersection of $i,z,Y$ bands. The effective area for this region is $5019 \ deg^{2}$. In brown shaded regions, we show the overlap between DES and VHS, covering $2,374 \ deg^{2}$ with $J,K_{s}$ and $1331 \ deg^{2}$ if the $H$ band is included (in light brown).}
 \label{fig:y3footprint}
\end{figure}

We use the $DES \cap VHS \cap AllWISE$ sample (called target sample) to find BDs. Matching the three catalogues and removing sources that do not pass the DES quality cuts, we find 42,046,583 sources. Applying the cut in signal-to-noise (SNR) greater than $5\sigma$ in $z,Y,J$ and selecting sources with $z<22$, we find 27,249,118 sources in our $2,374 \ deg^{2}$ footprint. 

We do not account for interstellar reddening. Our target sample is concentrated in the solar neighbourhood and therefore, applying any known extinction maps, like ~\citet{Sch98} or ~\citet{sch11} maps, we would overestimate the reddening.

\begin{table}
\centering
\caption{Information about the photometric passbands from DES Y3, VHS DR3 and AllWISE. Columns are the survey acronym, the filter name, the effective wavelength and the magnitude limit at $5 \sigma$ in AB.}
\label{tab:errcurve}

\begin{tabular}{c | c | c | c | }
    
    \hline \hline
    
Survey & Filter & $\lambda_{c}$ & $m_{5\sigma}$  \\ & & ($\mu m$) & (AB)  \\ \hline \hline
                              
DES & $i$   & 0.775 & 23.75   \\ \hline
DES & $z$   & 0.925 & 23.05   \\ \hline
DES & $Y$   & 1.0 & 20.75   \\ \hline
VHS & $J$   & 1.25  & 21.2   \\ \hline
VHS & $H$   & 1.65  & 19.85   \\ \hline
VHS & $K_{s}$  & 2.15 & 20.4   \\ \hline
AllWISE & $W1$  & 3.4 & 19.80   \\ \hline
AllWISE & $W2$  & 4.6 & 19.04 \\ \hline

\hline
    \end{tabular}
   \end{table}

\section{Calibration samples}
\label{sec:calib}


In this section, we present and characterise the calibration samples used to define our colour selection (Section~\ref{sec:targetselection}) and to create the photometric templates (see Section~\ref{subsec:calibproperties}) that will be used during the spectral classification step (Section~\ref{sec:classif_y3}) and to feed the \textit{GalmodBD} simulation (Section~\ref{sec:simulation}). Quasars are only used as a reference in the first stage, while the M dwarfs and BDs are used both to select the colour space and to create the colour templates that will be used during the analysis.

The calibration samples are: J. Gagn\'e's compilation of brown dwarfs, M dwarf sample from SDSS and spectroscopically confirmed quasars from DES. It is important to note that all of them have spectroscopic confirmation. 

Each calibration sample has been matched to the target sample. In all cases, we again use a matching radius of $2\arcsec$ to DES coordinates.

\subsection{Gagn\'e's sample of brown dwarfs}
\label{subsec:gagne}


The Gagn\'e compilation \footnote{\url{https://jgagneastro.wordpress.com/list-of-ultracool-dwarfs/}} contains a list of most of the spectroscopically confirmed BDs up to 2014, covering spectral types from late M to LT dwarfs. It consists of 1772 sources, covering most parts of the sky to distances less than $100 \ pc$. The spectral classification in this sample is given by its optical classification anchored to the standard L~dwarf scheme of ~\citet{kir99} or by its near-infrared classification anchored to the scheme for T~dwarfs from ~\citet{bur06b}. Some of the brown dwarfs present in this sample have both classifications. In most cases, both estimations agree, but for a few of them ($\approx 10\%$), there is a discrepancy of more than one spectral types, which can be considered as due to peculiarities. In these cases, we adopted the optical classification. We tested the effect of using one or another value in the creation of the templates and found discrepancies of $\lesssim 3\%$. 

From the initial list of 1772 BDs, we removed objects that are considered peculiar (in colour space) or that are part of a double system, categories given in the Gagn\'e compilation and also sources with spectral type M, yielding a remaining list of 1629 BDs. From this list, 233 are present in the DES footprint, but when we match at $2\arcsec$ between the DES and Gagn\'e sample, we only recover 150 of these. For the remaining 83 that are not matched within $2\arcsec$, we visually inspected the DES images to find their counterparts beyond the $2\arcsec$ radius, recovering in this process 58 additional LT types to the DES sample. Therefore, our sample totals 208 known LT dwarfs, while 25 are not found due to partial coverage of the footprint or due to high proper motions. 

Repeating the same procedure but for VHS: in the entire VHS footprint (not only in the $DES \cap VHS$ region), but we find 163 LT dwarfs at $2\arcsec$ radius. Here we did not repeat the process of manually recovering missing objects. In the $DES \cap VHS$ region, we end with 104 confirmed LT dwarfs, from 139 in the $DES \cap VHS$ common footprint. The missing sources are due to the same effects of partial coverage or high proper motions.

During our analysis, we found that BDs tagged as ``young" in the Gagn\'e sample were biasing the empirical colour templates used for classification (see details in Section~\ref{sec:classifcalib}).  BDs are tagged as ``young" in the Gagn\'e sample whenever they are found as members of a Young Moving Group (YMG) or are otherwise suspected of having ages less than a few $100 \ Myrs$. Young BDs are typically found to exhibit redder photometric colours in the near-infrared due to the effects of low-gravity and/or different cloud properties \citep[e.g.][]{faherty2016}. We, therefore, removed those from the calibration sample. 

As a result, our final LT calibration sample contains 208 sources in the Y3 DES sample, 104 in the $DES \cap VHS$ region, 163 in VHS DR3 alone and 128 with $VHS \cap AllWISE$. These are the final samples we use to calibrate the empirical colour templates for BDs, depending on the colour to parameterize (For example, for $(J-H)$ we use the VHS sample with 163 sources, whereas for $(i-z)$ and $(z-Y)$, we use the DES sample with 208 sources). These templates will also feed the \textit{GalmodBD} simulation for the LT population (Section \ref{sec:simulation}).

\begin{figure}
 \includegraphics[width=\columnwidth]{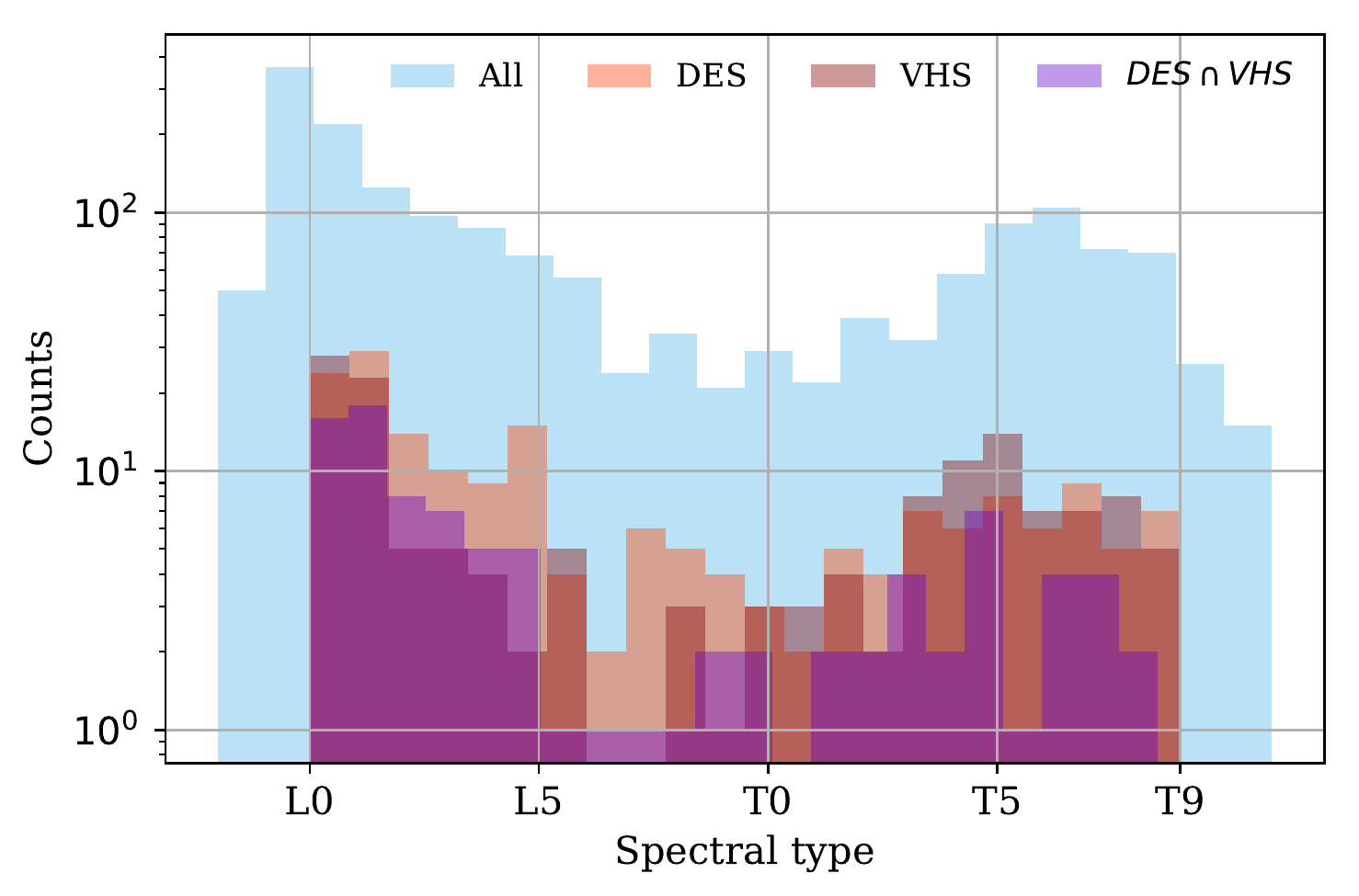}
 \caption{The distribution of known BDs in the Gagn\'e sample for different overlaps. The original sample contains 1,629 BDs, 208 of which are found in the DES Y3 data, 163 in the VHS DR3 data and 104 in the $DES \cap VHS$ data, after removal of M types and young L types.}
  \label{fig:calibGagne}
\end{figure}

In Fig.~\ref{fig:calibGagne} we show the number of BDs as a function of the spectral type in the Gagn\'e sample, matched to different photometric data. At this point, we assume that the colour templates we will obtain from these samples are representative of the whole brown dwarf population. Later, in section \ref{sec:targetselection}, we will compare the target sample to the calibration sample and confirm that this approximation is valid.

\subsection{M dwarfs}
\label{subsec:mstars}

The sample of M dwarfs comes from a spectroscopic catalogue of 70,841 visually inspected M dwarfs from the seventh data release of the SDSS \citep{west11}, confined to the Stripe82 region. After matching with the target sample at a $2\arcsec$ radius, we end up with 3,849 spectroscopically confirmed M dwarfs with spectroscopic classification, from M0 to M9. We use this sample to create templates for our classification schema, with particular care for the transition from M9 to L0. These templates will also feed the \textit{GalmodBD} simulation for the M population (Section \ref{sec:simulation}).

\subsection{Quasars}
\label{subsec:qso}

Quasars have been traditionally a source of contaminants in brown dwarf searches since they are point-like, and at high redshift, they can be very red, especially in infrared searches using WISE or 2MASS data, for instance. On the other hand, this degeneracy can be broken with the use of optical information, as shown in ~\citet{ree15} and ~\citet{ree17}. We use two samples of confirmed quasars in DES, one from ~\citet{tie17} up to z=4 and the one from ~\citet{ree17} with $z>6$. In principle, quasars follow a different colour locus (as seen in~\citet{ree17}) but some contamination might remain after the colour cuts, and we will treat them as a source of extragalactic contamination in Section~\ref{sec:sg}.

\subsection{Colour templates}
\label{subsec:calibproperties}

To define our colour selection, to classify BDs and to produce realistic colours in the simulation, we create colour templates as a function of the spectral type in $(i-z)_{AB}$, $(z-Y)_{AB}$, $(Y_{AB}-J_{Vega})$, $(J-H)_{Vega}$, $(H-K_{s})_{Vega}$, $(K_{s}-W1)_{Vega}$ and $(W1-W2)_{Vega}$. 

Since M dwarfs are much more abundant than BDs in our samples, we adopt different approaches to build the colour templates, depending on the available number of calibrating sources. For M and L0 dwarfs, where we have enough statistics, we take the mean value for each spectral type as the template value, selecting sources with $SNR > 5\sigma$ only. Beyond L0, since we do not have enough statistics for all spectral types, we follow a different approach: we fit the colour versus spectral type distribution locally in each spectral type using both first order and second order polynomials. For instance, for L7 we fit the colour distribution between L3 and T3, and with the given first and second order polynomials, we interpolate the result for L7. Finally, the colour value for the given spectral type is taken as the average of the two polynomial fits. 

\begin{figure*}
\begin{center}
    \includegraphics[width=0.95\linewidth]{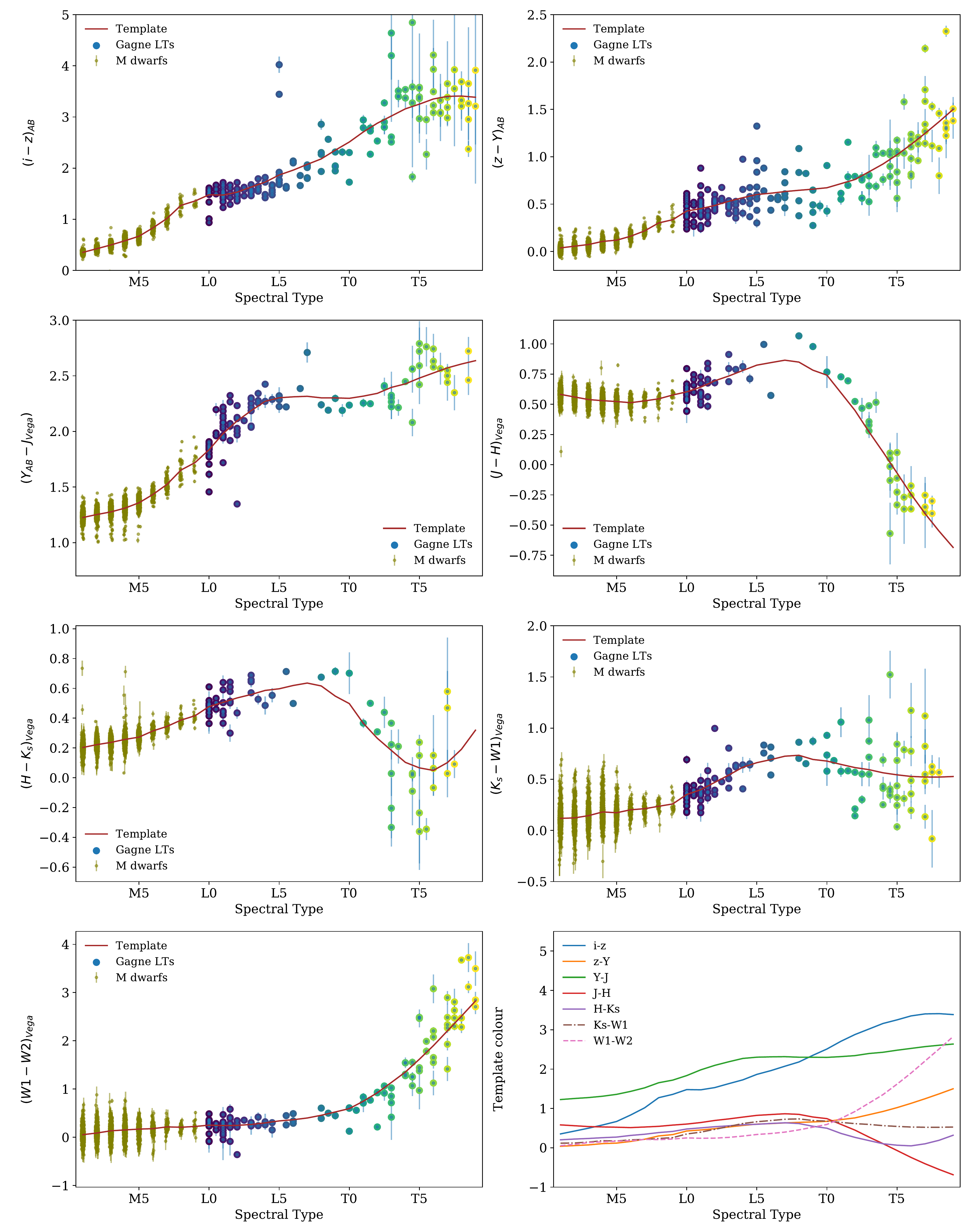}
\end{center}
\caption{Colours as a function of the spectral type in the MLT spectral regime. M dwarfs are shown in olive, in blue, green and yellow are BDs with their colour given by their spectral type, as seen in Fig~\ref{fig:datacolor}. These templates are used during the classification and to feed the \textit{GalmodBD} simulations. In the last panel, we compare all the templates together. Since several colours exhibit degeneracies in their colour-spectral type relationships, it is essential to have multiple colours to establish a good spectral type calibration.}
\label{fig:sptcolor}
\end{figure*}

The empirical templates can be seen in Fig.~\ref{fig:sptcolor}. The template values are listed in Table~\ref{tab:galmodcolors}. We found several degeneracies in colour space. For example, in $(H-K_{s})$, one cannot distinguish spectral types between early-L and late-L/early-T because their colours are the same, or in $(W1-W2)$, where we cannot discriminate between mid-M stars and early-mid/L-dwarfs for the same reason. Since several colours exhibit degeneracies in their colour-spectral type relationships, it is important to have multiple colours to establish a good spectral type calibration and therefore the need for a combination of optical and infrared data.

In terms of the dispersion about the templates, it increases with spectral type, with some exceptions. For example, in $(W1-W2)$ the dispersion for M types is larger than for some LT types. In the T regime, in general, the dispersion due to variations in metallicity, surface gravity or cloud cover, among other effects, can be of the same order or bigger than the dispersion introduced by differences in signal-to-noise. Possible peculiar objects will be identified with high $\chi^2$ when compared to the empirical templates.

In the last panel of Fig.~\ref{fig:sptcolor}, we compare all the templates together. From here, we find that $(i-z)$ colour has the largest variation through the ML range, demonstrating the importance of the optical filters to separate M dwarfs from LT dwarfs. $(W1-W2)$ is very sensitive to T types (as expected by design), and $(Y-J)$ is also important for the ML transition. The other bands will add little to the M/L transition, but they will help on L/T transition. Similar findings were already presented in ~\citet{skr14}.

\begin{table*}
\centering
\caption{Template values as a function of the spectral type in the MLT regime. These are the best-fit values shown in Fig.~\ref{fig:sptcolor} and that are later used in the classification and to feed the \textit{GalmodBD} simulation.} 
\label{tab:galmodcolors}

\begin{tabular}{| c | c c c c c c c |}
        \hline \hline
    
Spectral Type & & &  & Colour & & &  \\
 \hline
 & $(i-z)_{AB}$ & $(z-Y)_{AB}$ & $(Y_{AB}-J_{Vega})$ & $(J-H)_{Vega}$ & $(H-K_{s})_{Vega}$ & $(K_{s}-W1)_{Vega}$ & $(W1-W2)_{Vega}$
 \\ \hline

M1 & 0.35 & 0.04 & 1.23 & 0.58 & 0.20 & 0.12 & 0.05 \\
M2 & 0.43 & 0.06 & 1.25 & 0.56 & 0.22 & 0.12 & 0.09 \\
M3 & 0.50 & 0.07 & 1.28 & 0.54 & 0.24 & 0.14 & 0.13 \\
M4 & 0.58 & 0.11 & 1.31 & 0.53 & 0.26 & 0.18 & 0.15 \\
M5 & 0.67 & 0.12 & 1.36 & 0.52 & 0.27 & 0.17 & 0.17 \\
M6 & 0.83 & 0.16 & 1.43 & 0.51 & 0.31 & 0.20 & 0.18 \\
M7 & 1.02 & 0.22 & 1.52 & 0.53 & 0.34 & 0.21 & 0.21 \\
M8 & 1.27 & 0.30 & 1.65 & 0.54 & 0.39 & 0.24 & 0.21 \\
M9 & 1.36 & 0.34 & 1.72 & 0.58 & 0.42 & 0.26 & 0.22 \\
\hline
L0 & 1.48 & 0.43 & 1.83 & 0.60 & 0.48 & 0.35 & 0.25 \\
L1 & 1.47 & 0.45 & 1.98 & 0.64 & 0.51 & 0.40 & 0.24 \\
L2 & 1.53 & 0.49 & 2.09 & 0.69 & 0.54 & 0.47 & 0.25 \\
L3 & 1.63 & 0.53 & 2.19 & 0.73 & 0.56 & 0.54 & 0.26 \\
L4 & 1.73 & 0.57 & 2.27 & 0.78 & 0.59 & 0.62 & 0.30 \\
L5 & 1.87 & 0.60 & 2.30 & 0.82 & 0.60 & 0.66 & 0.34 \\
L6 & 1.96 & 0.62 & 2.31 & 0.84 & 0.62 & 0.69 & 0.36 \\
L7 & 2.07 & 0.63 & 2.32 & 0.87 & 0.64 & 0.72 & 0.40 \\
L8 & 2.18 & 0.65 & 2.30 & 0.85 & 0.62 & 0.73 & 0.46 \\
L9 & 2.35 & 0.66 & 2.30 & 0.78 & 0.55 & 0.69 & 0.52 \\
\hline
T0 & 2.51 & 0.67 & 2.30 & 0.74 & 0.50 & 0.68 & 0.59 \\
T1 & 2.71 & 0.72 & 2.32 & 0.60 & 0.36 & 0.64 & 0.75 \\
T2 & 2.88 & 0.76 & 2.34 & 0.45 & 0.26 & 0.61 & 0.92 \\
T3 & 3.02 & 0.84 & 2.40 & 0.27 & 0.18 & 0.59 & 1.13 \\
T4 & 3.16 & 0.92 & 2.43 & 0.11 & 0.10 & 0.56 & 1.35 \\
T5 & 3.25 & 1.02 & 2.48 & -0.07 & 0.07 & 0.54 & 1.62 \\
T6 & 3.35 & 1.13 & 2.53 & -0.24 & 0.05 & 0.53 & 1.90 \\
T7 & 3.40 & 1.25 & 2.57 & -0.40 & 0.10 & 0.52 & 2.21 \\
T8 & 3.41 & 1.37 & 2.61 & -0.55 & 0.20 & 0.52 & 2.51 \\
T9 & 3.39 & 1.50 & 2.64 & -0.68 & 0.32 & 0.53 & 2.83 \\ 

\end{tabular}
\end{table*}

\section{MLT colour selection}
\label{sec:targetselection}

\begin{figure*}
\begin{center}
    \includegraphics[width=\linewidth]{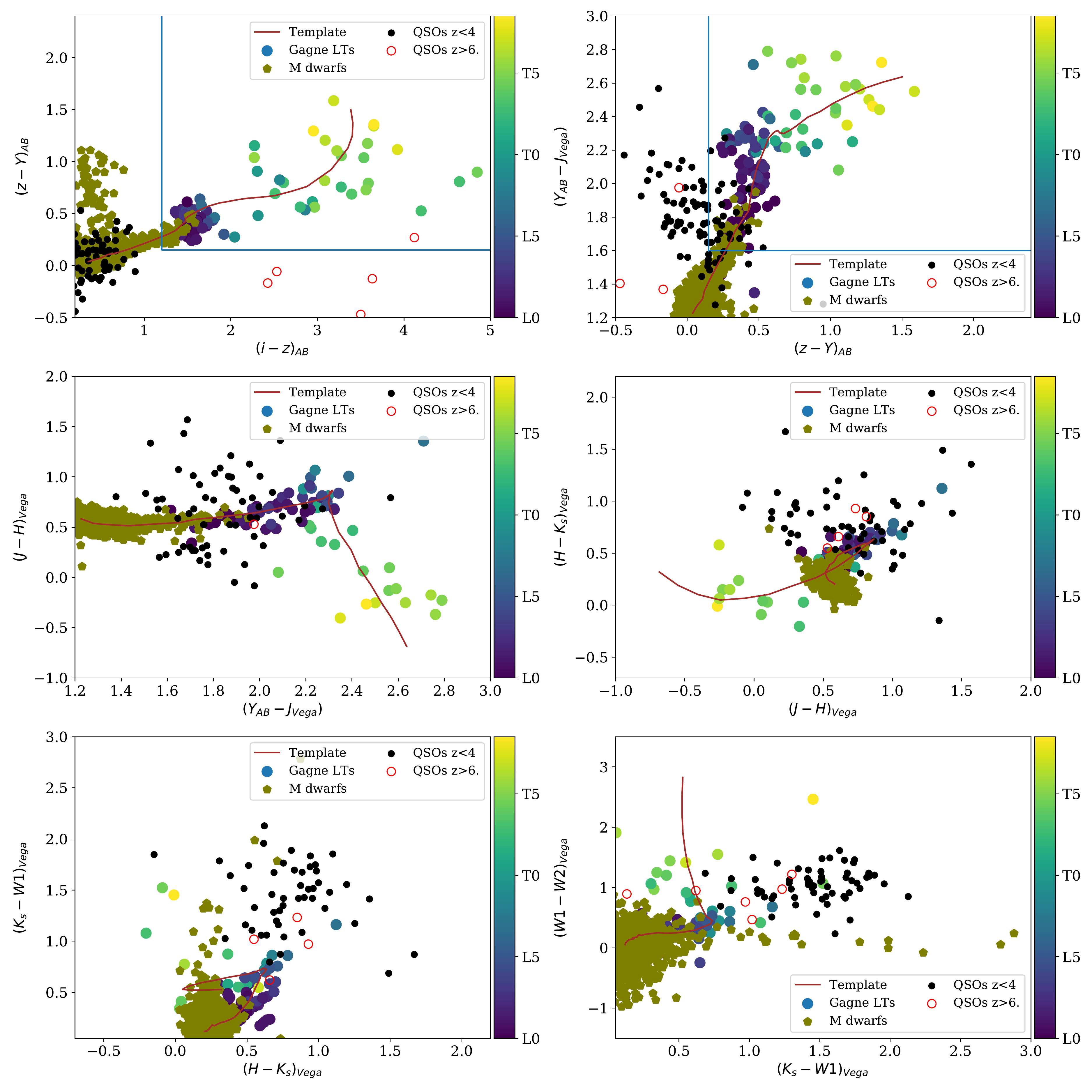}
\end{center}
\caption{Colour-colour diagrams for spectroscopically confirmed BDs, M dwarfs and quasars that are found in the target sample. Each panel corresponds to a specific colour-colour space. In all panels, M dwarfs are shown as olive circles. Quasars as black circles for $z<4$ and as empty red circles for $z>6$, whereas LT dwarfs are shown as filled circles with their colour a function of the spectral type, as depicted in the colour bar on the right. The colour templates that were empirically derived (see Section \ref{subsec:calibproperties}) are shown by the brown line. The colour cuts applied to the sample before classification are depicted by the blue lines in the first two panels.}
\label{fig:datacolor}
\end{figure*}

In this section we explain the steps to select our initial list of LT candidates from the target sample. In Fig.~\ref{fig:datacolor}, we show the colour-colour diagrams of known BDs, M dwarfs and quasars. Clearly, some of these colour-colour diagrams are more efficient to disentangle LT dwarfs from other point sources than others. Furthermore, at $z \sim 22$, we are still mostly complete in the $i$ band, although not necessarily for late T types. Therefore we do not demand detection in $i$ band, but we still impose a minimum $(i-z)$ colour as a selection criterion, very efficient to remove quasars from our sample, as attested by Fig.~\ref{fig:datacolor}. 

The nominal cut at $z_{AB} = 22$ is set to ensure that the combination $DES \cap VHS$ takes full advantage of both surveys, something that was not possible earlier due to the brighter magnitude limits imposed by SDSS. At $z_{AB} = 22$, we are limited at $J_{Vega} \sim 19.7$. This corresponds to $z_{AB}-J_{Vega} \sim 2.30$, which is the colour of an L0 according to the templates presented in Table~\ref{tab:galmodcolors}. This is  brighter by at least $0.7$ magnitudes than the $5\sigma$ limit on VHS. In other words, we will be able to reach $z_{AB}=22.7$ in upcoming DES updates and still remain complete in optical and VHS bands for the LT types. The magnitude limits on VHS bands are also deeper than its predecessors. For example, UKIDSS has a global depth limit of $J_{Vega} \sim 19.6$ \citep{war07}, while VHS has $J_{Vega} \sim 20.3$. 

We define the colour space where BDs are found to reside. We initially aim at high completeness in color space, at the expense of allowing for some contamination by M dwarfs and extragalactic sources. Purity is later improved at the classification stage (Section \ref{sec:classif_y3}). 

The entire selection process can be summarized in 7 stages as detailed below and listed in Table~\ref{tab:colorcuts}:

\begin{enumerate}

\item Quality cuts on DES and matching to VHS (explained in Section \ref{sec:data}). We end up with 42,046,583 sources after applying a matching within $2\arcsec$ between the DES and the VHS, and selecting sources with \verb+SEXTRACTOR_FLAGS_z,Y = 0+ and \verb+IMAFLAGS_ISO_i,z,Y = 0+. 

\item Magnitude limit cut in $z<22$ and signal-to-noise greater than $5\sigma$ in $z,Y,J$. We end up with 28,259,901 sources. 

\item We first apply a cut in the optical bands $(i-z)$, removing most quasar contamination while maintaining those sources for which $i$ band has no detection. We decided to apply a cut of $(i-z)>1.2$. In comparison with previous surveys, as seen in Table~\ref{tab:prevsurveys}, this is a more relaxed cut, initially focusing on high completeness in our sample at the expense of purity. This cut eliminates more than 99.8\% of the catalogue, leaving us with a sample of 65,041 candidates. 

\item We apply a second selection to the target sample in the space $(z-Y)$ vs. $(Y-J)$. From Fig.~\ref{fig:datacolor}, we decided the cut to be: $(z-Y) > 0.15$ , $(Y-J) > 1.6$ avoiding the $z>6$ QSO colour locus. The surviving number of sources is 35,548. This cut is actually very similar to the cut proposed in \citet{skr2016}. They imposed $Y-J > 0.8$ (Vega). Transforming our $Y$ band to Vega, our equivalent cut would be $Y-J > 0.7$. 
\item Finally, we apply the footprint mask. In this process, we end up with 35,426 candidates. This is the sample that goes into classification. Extragalactic contamination is treated after running the classification.

\item Apply the \textit{photo-type} method \citep{skr14} to estimate the spectral type in the MLT range for all the targets. We call this method \textit{classif} (see Section~\ref{sec:classif_y3}). After removal of the M types, we ended with 12,797 LT types.

\item Removal of extragalactic contamination (see Section~\ref{sec:sg}). We end up with the final sample of 11,745 LT types from which 11,545 are L types, and 200 are T types. There is also extragalactic contamination in the M regime, as explained in the next section.


\end{enumerate}

\begin{table*}
\centering
    \caption{Steps used in this paper to classify LT sources in $DES \cap VHS \cap WISE$. First, quality cuts are applied to the data to remove spurious targets. Next, a magnitude limit is imposed in the $z$ band and finally, colour cuts are applied to select only the reddest objects. These are the sources that enter the classification. Finally, extragalactic contamination is removed.}
    \label{tab:colorcuts}
    \begin{tabular}{c c c c}
    \hline 
   Step & Description & Percentage &  Number  \\ 
        &             & Removed&  Remaining \\
\hline \hline
   0    & DES Y3 sample (DR1) & & 399,263,026 \\
   \hline
   1    & Matching 2 arcseconds to VHS & &  \\
        & \verb+FLAGS_z,Y=0+ && \\
        & \verb+IMAFLAGS_ISO_i,z,Y=0+ & 89.5\%  & 42,046,583 \\
        \hline
   2    & $z<22$ && \\
        & $SNR\_{z,Y,J} > 5\sigma$ & 33\% & 28,259,901 \\
        \hline
   3    & $(i-z)_{AB} > 1.2$ & 99.8\% & 65,041 \\ 
   \hline
   4    & $(z-Y)_{AB} > 0.15$ && \\
        & $Y_{AB}-J_{Vega} > 1.6$ & 45\% & 35,548 \\ 
        \hline
   5    & Footprint masking &   0.3\%    & 35,426\\ 
   \hline
   6    & LT Classification & 64\% & 12,797 \\ 
   \hline
   7    & Remove extragalactic contamination & 8\% & 11,745 \\  
\hline
    \end{tabular}
    
      \end{table*}

\section{Spectral classification}
\label{sec:classif_y3}

In this section we explain the method to assign spectral types to our candidate list and evaluate its uncertainties based on the calibration samples. We implement the same classification method presented in \citet{skr14} and \citet{skr2016}, based on a minimization of the $\chi^2$ relative to the MLT empirical templates. Our method uses the templates created in Section~\ref{subsec:calibproperties}. We refer to our classification code as \textit{classif}.  

As mentioned earlier, we impose a $5\sigma$ detection in $z,Y,J$. For the rest of the bands, we require a $3\sigma$ detection. If the magnitude error exceeds this limit, we consider the source as not observed in the given band. However, there is one exception, the $W1$ magnitude, which is cut at $5\sigma$ in the AllWISE catalogue.

Let a set of $N_{bands}$ observed for the j-th candidate to be \{$m_{ji}$, i=1,$N_{bands}$\} and their photometric uncertainties be  \{$\sigma_{m_{ji}}$, i=1,$N_{bands}$\}. Let also a set of colour templates be \{$c_{i,k}$, i=1,$N_{bands}$, k=1,$N_{mod}$\}, which give the magnitude difference between band $b$ and the reference band $z$. Hence $c_{z,k}=0$ by construction. Let's consider the template intrinsic dispersion to be \{$\sigma_{c_{i,k}}$, i=1,$N_{bands}$, k=1,$N_{mod}$\}, which we fix for all bands and templates to $\sigma_{c} = 0.07$, as also done by \citet{skr2016}. The total error for band b, for the j-th candidate will be $\sigma_{bj} = \sqrt{\sigma_{m_{jb}}^{2}+\sigma_{c}^{2}}$.

The first step in the classification process for the j-th candidate is, for each of the k-th spectral type ($N_{mod}$), calculate the inverse variance weighted estimate of the reference magnitude (in our case z) as:

\begin{equation}
\hat{m}_{j,z,k}=\frac{\sum_{b=1}^{N_{bands}} {\frac{m_{jb}-c_{bk}}{\sigma_{bj}^{2}}}} {\sum_{b=1}^{N_{bands}} {\frac{1}{\sigma_{bj}^{2}}}}
\label{eq:mBt}
\end{equation}

Next, the above value is used to calculate the $\chi^{2}$ value for the k-th spectral type, for each j-th candidate:

\begin{equation}
\chi^2(\{m_{j}\},\{\sigma_{bj}\},\hat{m}_{z,k},k) = \sum_{b=1}^{N_{bands}}{(\frac{m_{jb}-\hat{m}_{z,k}-c_{b,k}}{\sigma_{bj}})}^2
\label{ec:chi2}
\end{equation}

Finally, we assign the spectral type that gives the minimum $\chi^{2}$ value.

\subsection{Classification performance on known samples}
\label{sec:classifcalib}

\begin{figure*}
\begin{multicols}{2}
    \includegraphics[width=\linewidth]{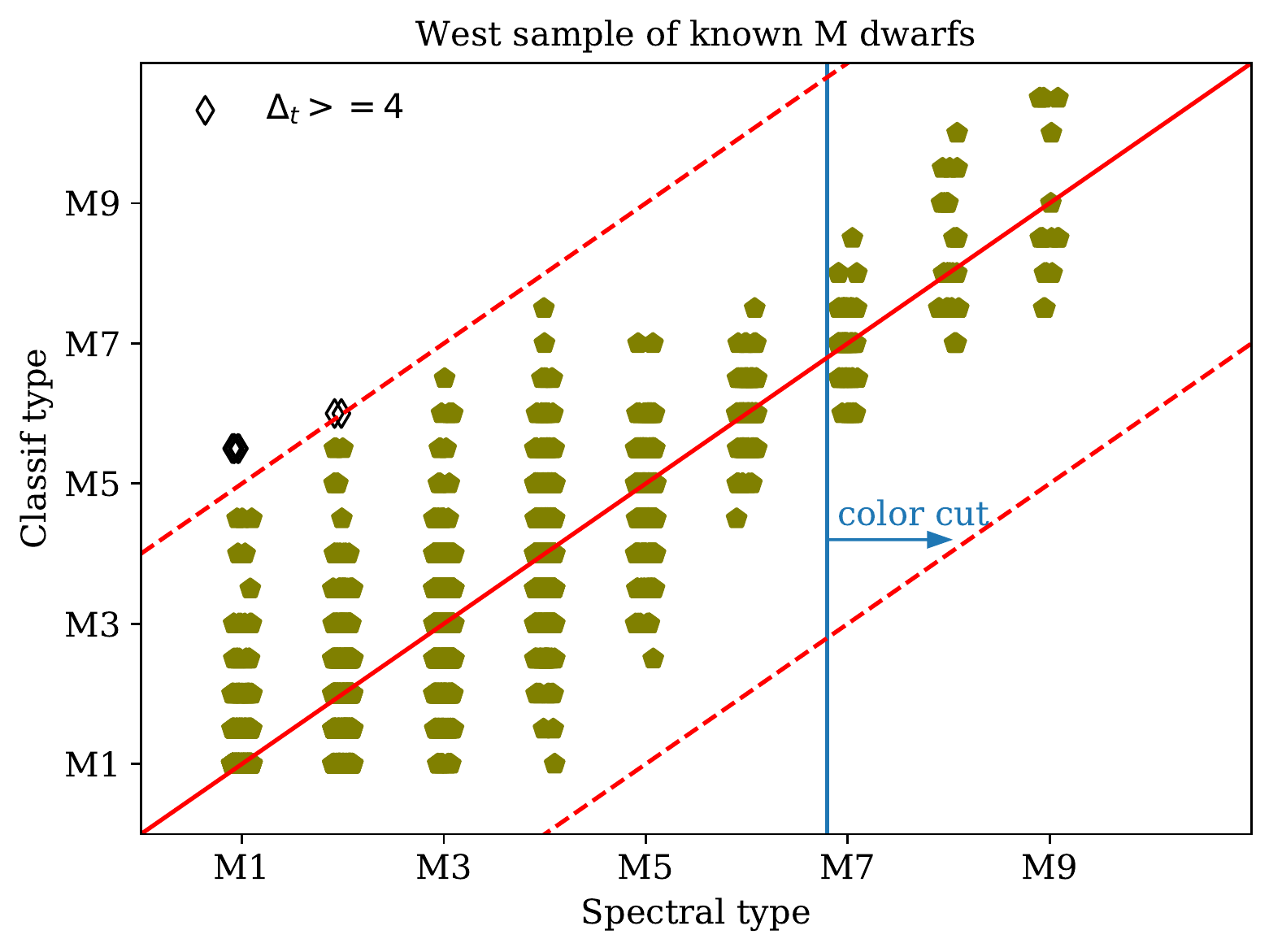}\par 
        \includegraphics[width=\linewidth]{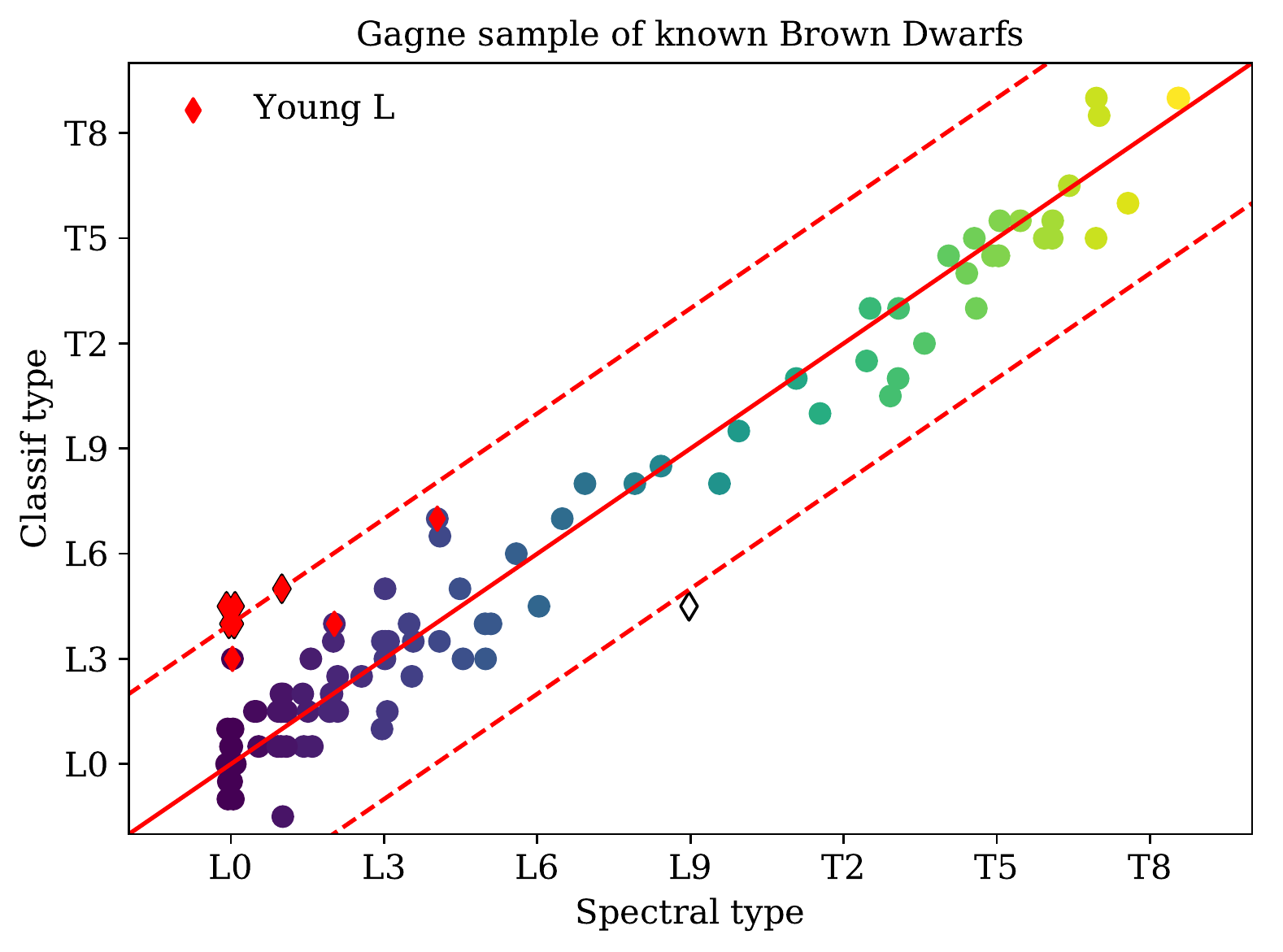}\par 
    \end{multicols}
\caption{Classification performance for M dwarfs (left) and LT types (right). In both panels, sources with a misclassification higher or equal 4 are shown as diamond symbols. Red dashed lines indicate the limits where the difference between the spectroscopic and photometric classification is equal to 4. On the right panel, we also tag young L types, which were not used to build the templates but are shown here to illustrate their peculiar nature, with a classification that is always higher than 2. The classification have a mean uncertainty of $\sigma_{M} = 0.69$ (for M dwarfs passing the colour cuts) and  $\sigma_{L} = 1.03$ and $\sigma_{T} = 1.12$ for the LT sample without the young L types. For visualization purposes we have applied small random shifts smaller than $\Delta_{template} \pm 0.1$ to the true spectral types (only in the x-axis).}
    \label{fig:classifcalib}

\end{figure*}

In this section we apply \textit{classif} to our list of known BDs and M dwarfs to assess how well the method works. We run the code on the Gagn\'e list of 104 known BDs (including young BDs) and in the West list of known M dwarfs. Results are summarized in Fig.~\ref{fig:classifcalib}, where we show the photometric spectral type versus the spectroscopic classification, separately for M (left) and LT types (right). In these figures, the diamond points are those where the difference between the true spectral type and the photometric estimation is greater or equal to 4 ($\Delta_{t} >= 4$). There are 8 sources with $\Delta_{t} >= 4$, from which 6 of them are young L types.

We estimate the accuracy of the method using: 

\begin{equation}
\sigma_{classif} = \frac{\sum_{j=1}^{N_{candidates}} {| \Delta t |}}{N_{candidates}} \frac{\sqrt{2N_{candidates}}}{2}
\label{eq:sigma}
\end{equation}

Dividing in M types (from West sample), L and T types (Gagn\'e sample), we get $\sigma_{M} = 0.69$ (for M with spectral type M7 or higher), $\sigma_{L} = 1.47$ and $\sigma_{T} = 1.12$ and a global $\sigma_{LT} = 1.37$. If we now estimate the errors but without the young L types, the metrics improve to $\sigma_{L} = 1.03$ and $\sigma_{T} = 1.12$ and a global $\sigma_{LT} = 1.06$, a precision compatible to what was found in \citet{skr2016}. Nonetheless, it is obvious from Fig.~\ref{fig:classifcalib} that it is much more appropriate to use a 3$\sigma$ value for how well one establish a photometric spectral type for these low-mass objects, thereby implying that the best one can do in estimating photometric spectral types is $\pm 2$ for M types and $\pm 3$ for LT-dwarfs.

\begin{figure}
    \includegraphics[width=\linewidth]{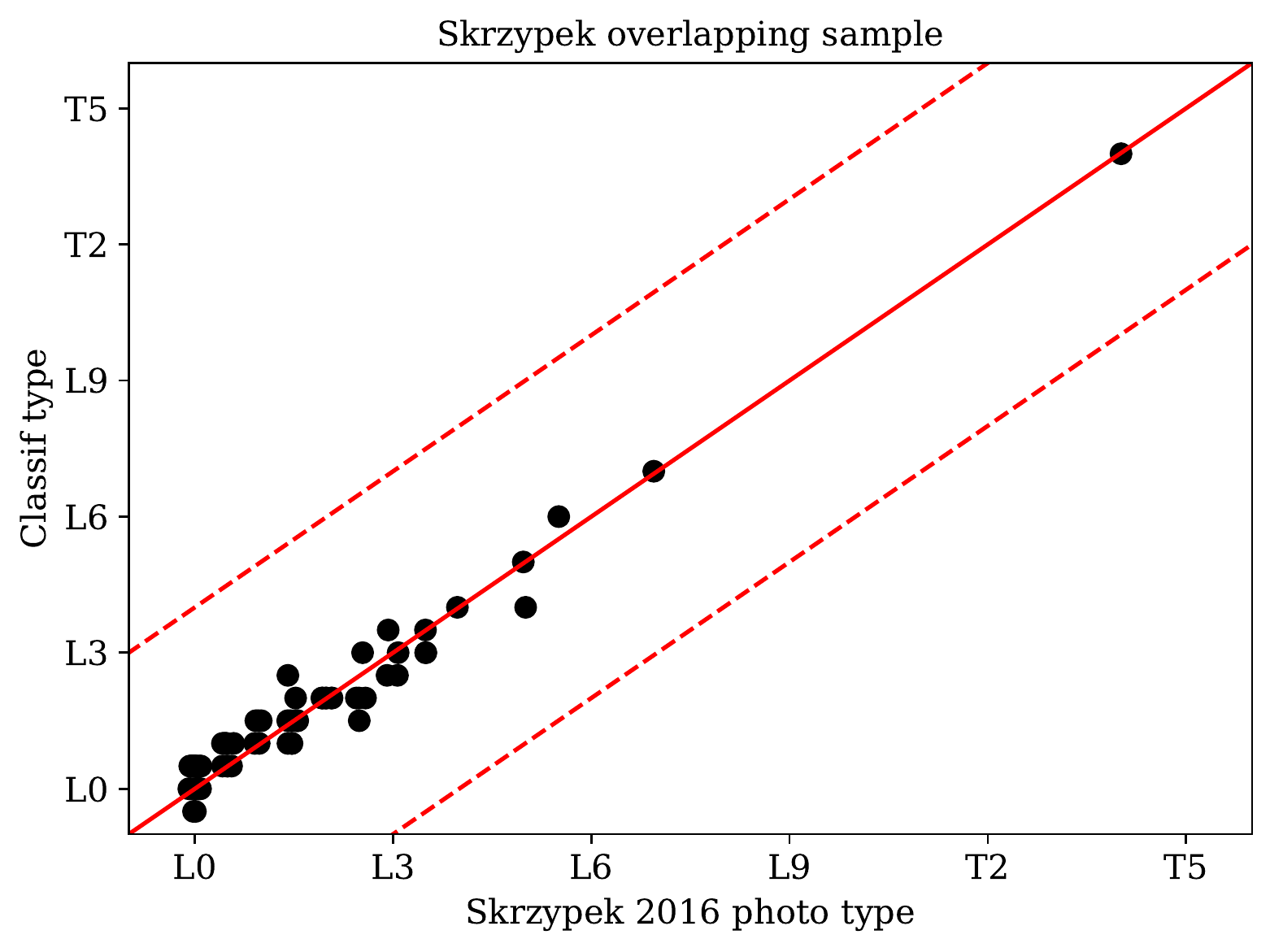}\par 
\caption{Comparison between the photometric classification of \citet{skr2016} and \textit{classif} for 74 common LT candidates found in the footprint. We find an excellent agreement between both estimates with a mean dispersion of $\sigma_{classif} = 0.38$. For visualization purposes we have applied small random shifts smaller than $\Delta_{template} \pm 0.1$ to the \citet{skr2016} types (only in the x-axis).}
        \label{fig:classifskr}

\end{figure}

Another test we perform is to estimate the spectral types for those LT candidates in the overlapping sample with \citet{skr2016}, consisting of 74 sources. We run \textit{classif} on these 74 sources and compare both photometric estimators. In general, we find an excellent agreement with $\sigma_{classif,skrzypek} = 0.38$. In Fig.~\ref{fig:classifskr} we show the comparison between the two methods.

\section{Classification of the target sample}
\label{mainclassif}

After confirming that our method reliably classifies MLT spectral types, we run \textit{classif} on the target sample presented in Section~\ref{sec:targetselection}, a sample of 35,426 candidates. A visual inspection of the candidates demonstrates that all are real sources. The main caveat in our methodology might be residual contamination by extragalactic sources. We next explain our star-galaxy separation method.

\subsection{Extragalactic contamination}
\label{sec:sg}

To remove possible extragalactic contamination, we run \verb+Lephare+ photo-z code on the whole candidate list using galaxy and quasar templates spanning various redshifts, spectral types and internal extinction coefficients (the \verb+Lephare+ configuration used is presented in Appendix~\ref{app:lephare}). For those candidates where the best-fit is $\chi_{lephare}^{2} < \chi_{classif}^{2}$, we assign them a galaxy or quasar class and are no longer considered MLT types. This method also has the potential to identify interesting extragalactic targets. 

It is worth noting that running \verb+Lephare+ onto the Gagn\'e sample, we can recover most known LT types. Only one brown dwarf in the Gagn\'e sample is assigned a galaxy class, which is already known to be a very peculiar L7 type called \textit{ULAS J222711-004547}~\citep{pec14}. It has a $\chi_{classif}^{2} > 630$, while the rest of the 103 BDs have a $\chi_{classif}^{2} < 130$. Also, the classification is robust concerning changes in the galaxy libraries used in the \verb+Lephare+ configuration: we tested various choices of galaxy templates, and the number of extragalactic contaminants remained constant within errors.

\subsection{Results}

After running \textit{classif} and \verb+Lephare+ on the target sample, we obtain 32,608 sources classified as MLT types. From these, 20,863 are M types, 11,545 are L types and are 200 T types. Our catalogue also includes 2,818 candidates classified as galaxies or quasars. A preliminary discussion about the properties of the extragalactic sources is given in Appendix~\ref{app:lepharesg}. In Section~\ref{app:catexplanation} we show an example of the data published in electronic format and its explanation. It consists of a table with all the 11,745 candidates in the target sample with LT spectral type including the photometry used and the \textit{classif} results. 

Sources with a number of bands available for classification (\verb+NBANDS+) less than 5 (3 or 4) are generally assigned to extragalactic spectral types; likewise those have the best-fit MLT template as M types. By visually inspecting the spectral energy distribution of sources with \verb+NBANDS+ < 5, compared to the best-fit templates of MLT types, galaxies and quasars, we conclude that the classification is ambiguous and should be taken with caution when \verb+NBANDS+ < 5. From the catalogue of 35,426 targets, 6\% have \verb+NBANDS+ < 5. If we consider only those with spectral type L, the percentage goes down to 3.7\% (469 targets), i.e., 3.7\% of the L types have \verb+NBANDS+ < 5, from which 96\% (449 targets) are assigned to a galaxy template instead of to an L type. This effect contributes to the uncertainty associated with the removal of extragalactic contamination. 

\begin{figure*}
\begin{center}
\includegraphics[width=0.97\textwidth]{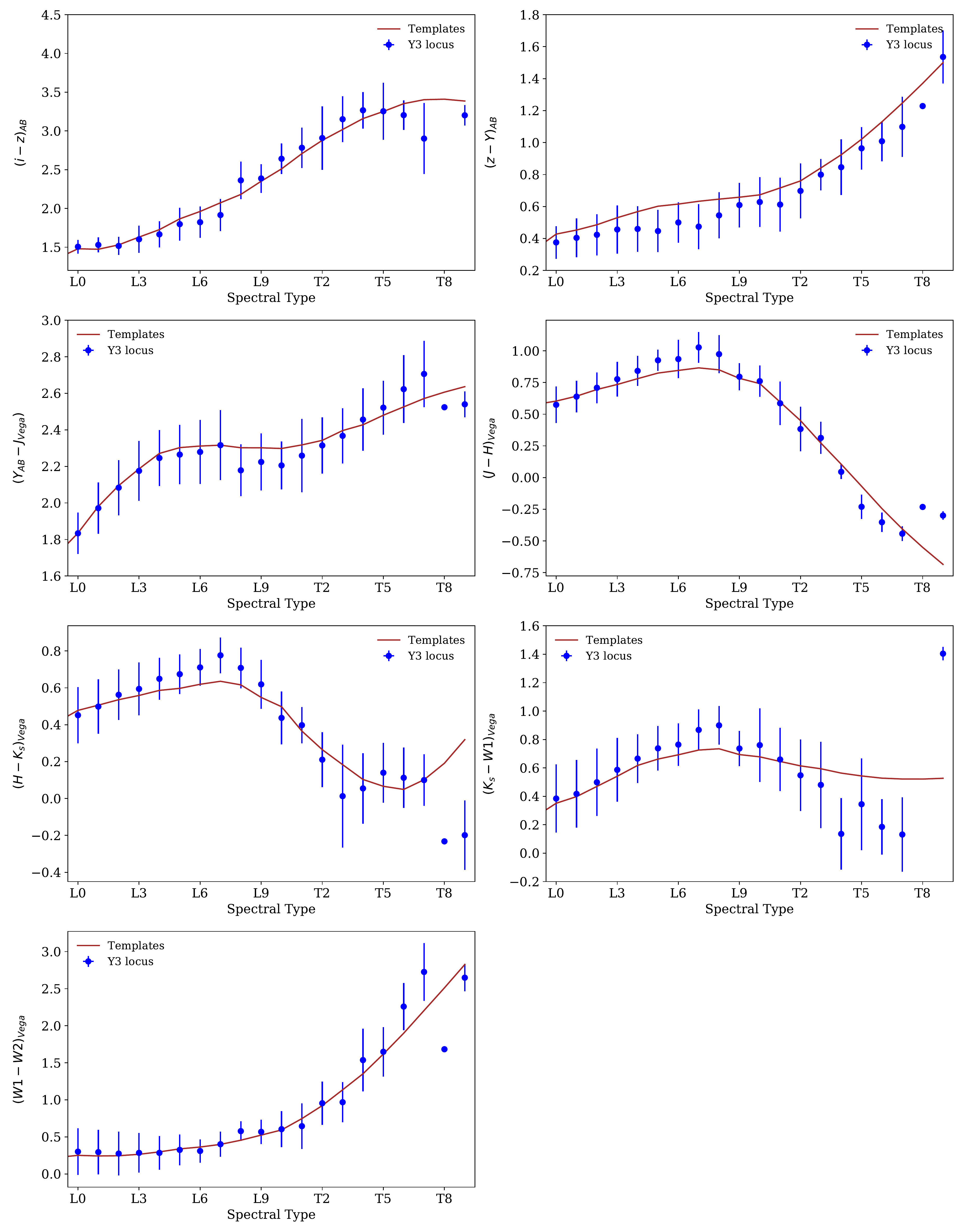}
\end{center}
\caption{Comparison between the colour templates and the locus of the LT candidates. Blue points show the mean and standard deviation colour for candidates with the given spectral type. In brown, the empirical templates used. The agreement is good through all the spectral space, with deviations appearing in the late T type regime, where the calibration sample is very sparse, confirming the initial assumption that the calibration sample is representative of the whole LT population.}
\label{fig:comparetemplates}
\end{figure*}

At this point, we compare the colour distribution as a function of the spectral type for the target sample with respect to the empirical templates of Section~\ref{subsec:calibproperties}. We assumed that the templates were representative of the LT population. In Fig.~\ref{fig:comparetemplates} we can see the comparison. Our modelling reproduces the target sample colour distribution. Only in the late T regime, we find some discrepancies. This effect is due to both the lack of statistics when we calculated the empirical templates in this regime, and to the intrinsic dispersion in colours in the T regime.

\subsection{Photometric properties of the LT population}
\label{subsec:photometricproperties}

\begin{figure}
    \includegraphics[width=\linewidth]{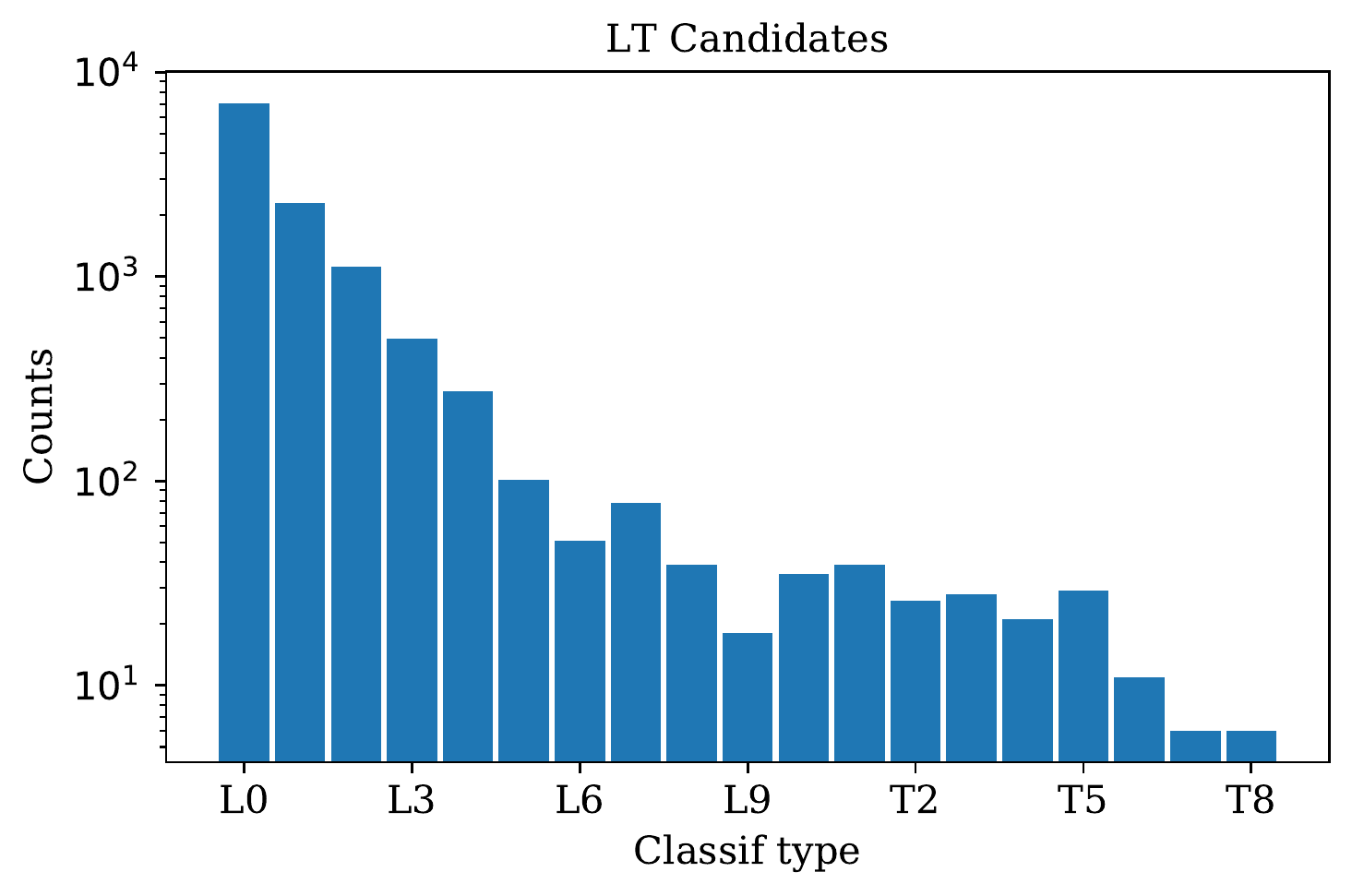}
\caption{The distribution of spectral types for the LT candidates. We classify 20,863 sources as M dwarfs (from M7 to M9, not shown here), 11,545 as L types (from L0 to L9) and 200 as T types (from T0 to T9).}
\label{fig:nzcandidates}

\end{figure}

\begin{figure}
    \includegraphics[width=\linewidth]{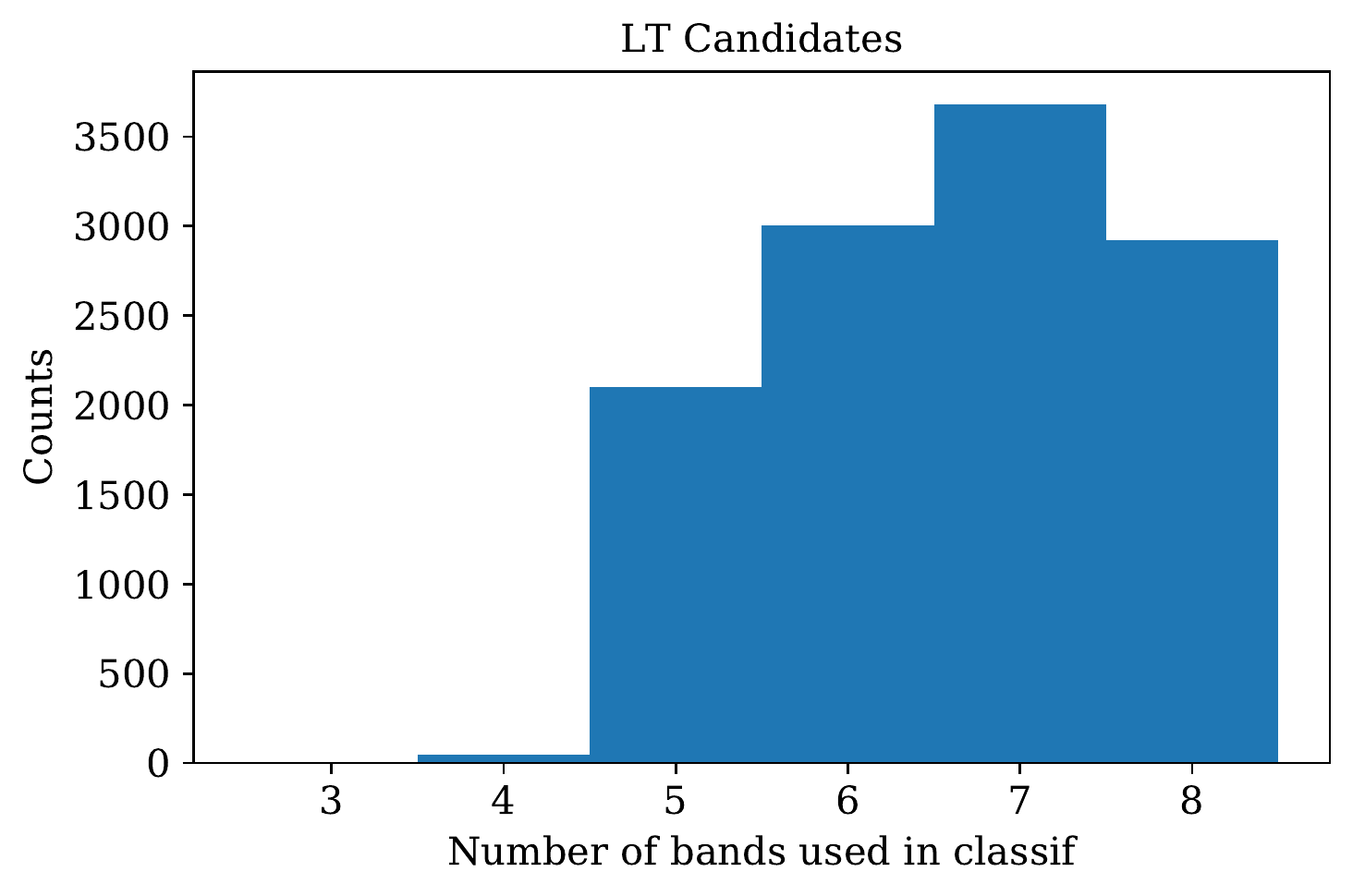}
\caption{The number of bands ($N_{bands}$) used during the classification. This information is used to assign weights when we calculate the theoretical $\chi^{2}$ distribution in Fig.~\ref{fig:chi2Lcandidates}.}
\label{fig:nbandscandidates} 

\end{figure}

We analyse the properties of the 11,745 LT types found in the target sample. The distribution of spectral types can be found in Fig.~\ref{fig:nzcandidates} in logarithmic scale. In Fig.~\ref{fig:nbandscandidates} we show the distribution of bands available for classification. 

\begin{figure}
\includegraphics[width=\linewidth]{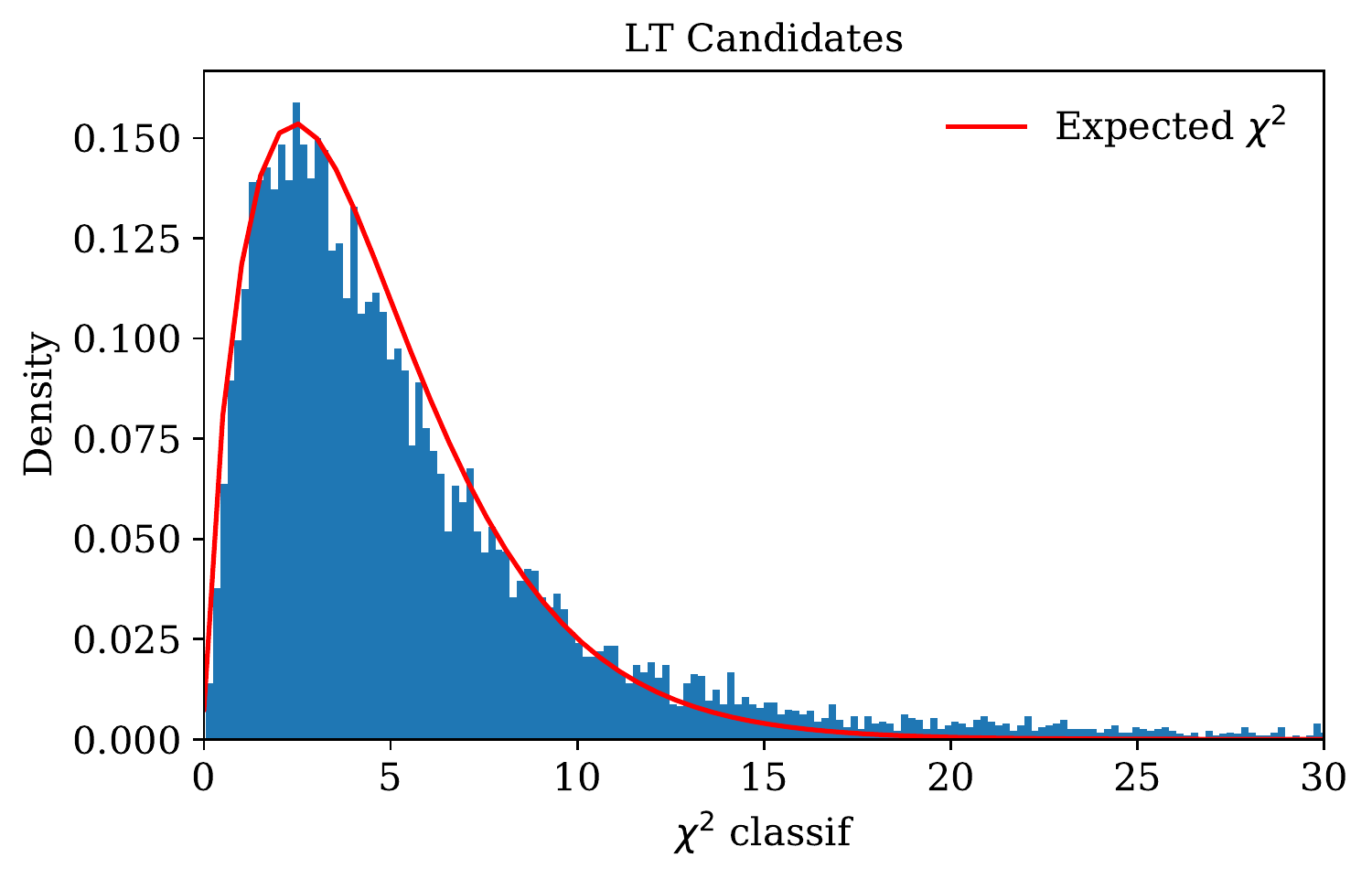}
\caption{$\chi^{2}$ distribution for the LT candidates, together with the theoretical expectation. In order to calculate this curve, we add, proportional to the numbers in Fig.~\ref{fig:nbandscandidates}, the degrees of freedom. Our distribution follows the expected curve, therefore our errors reflect the dispersion of the model. The reduced $\chi^{2}$ defined as $\frac{\chi^{2}}{(d.o.f)}$ is close to one, with a mean value of $\overline{\frac{\chi^{2}}{(d.o.f)}} = 1.3$ and a median value $med(\frac{\chi^{2}}{(d.o.f)}) = 0.95$.}
\label{fig:chi2Lcandidates}
\end{figure}

The $\chi_{classif}^{2}$ of the fits are generally very good, in agreement with the theoretical curve for the same degrees of freedom. In our model we have two free parameters, the brightness of the source and the spectral type, therefore the degrees of freedom are $N_{bands} - 2$. The $\chi_{classif}^{2}$ distribution can be found in Fig.~\ref{fig:chi2Lcandidates}. The theoretical curve is a summation over the $\chi^{2}$ curves for degrees of freedom $1,2,3,4,5$ (corresponding to $N_{bands} = 3,4,5,6,7,8$ respectively, where each curve was multiplied by the percentage of total sources with each number of bands. The reduced $\chi^{2}$ defined as $\frac{\chi^{2}}{(d.o.f)}$ is close to one, with a mean value of $\overline{\frac{\chi^{2}}{(d.o.f)}} = 1.3$ and a median value $med(\frac{\chi^{2}}{(d.o.f)}) = 0.95$.

In our analysis, we tag BDs as peculiar (in terms of colours) if their $\chi_{classif}^{2}$ is beyond 99.7\% of the probability, which in a z-score analysis means having a $\chi_{classif}^{2} > 3\sigma$ off the median. In our case, we set the cut-off at 3.5$\sigma$ instead to accommodate the natural dispersion of LT types. Since the $\chi_{classif}^{2}$ distribution is not normal we define $\sigma$ based on the median absolute deviation (MAD) instead of the standard deviation, and since the distribution is non-symmetrical, we apply a double MAD strategy. Since T types have intrinsically more colour dispersion than L types, we treat them separately. Applying the double MAD algorithm to the $\chi_{classif}^{2}$ distribution of L and T types, for a cut-off = 3.5$\sigma$, we end up with 461 L types labeled as peculiar ($\approx 4\%$) and 6 T types tagged as peculiar ($\approx 3.5\%$). These limits are equivalent to say that L types are peculiar whenever their $\chi_{classif}^{2}>20.$ and that T types are peculiar whenever their $\chi_{classif}^{2}>46.5$. Peculiar sources can be identified in the catalogue reading the \verb+PECULIAR+ column as explained in Section~\ref{app:catexplanation}.      

\subsubsection{Photometric distances}

We estimate photometric distances using the distance modulus: 

\begin{equation}
d(Type)[i] = 10*10^{(m_{z}[i]-M_{z}(Type))/5} 
\label{eq:distance}
\end{equation}

\begin{figure}
\includegraphics[width=\linewidth]{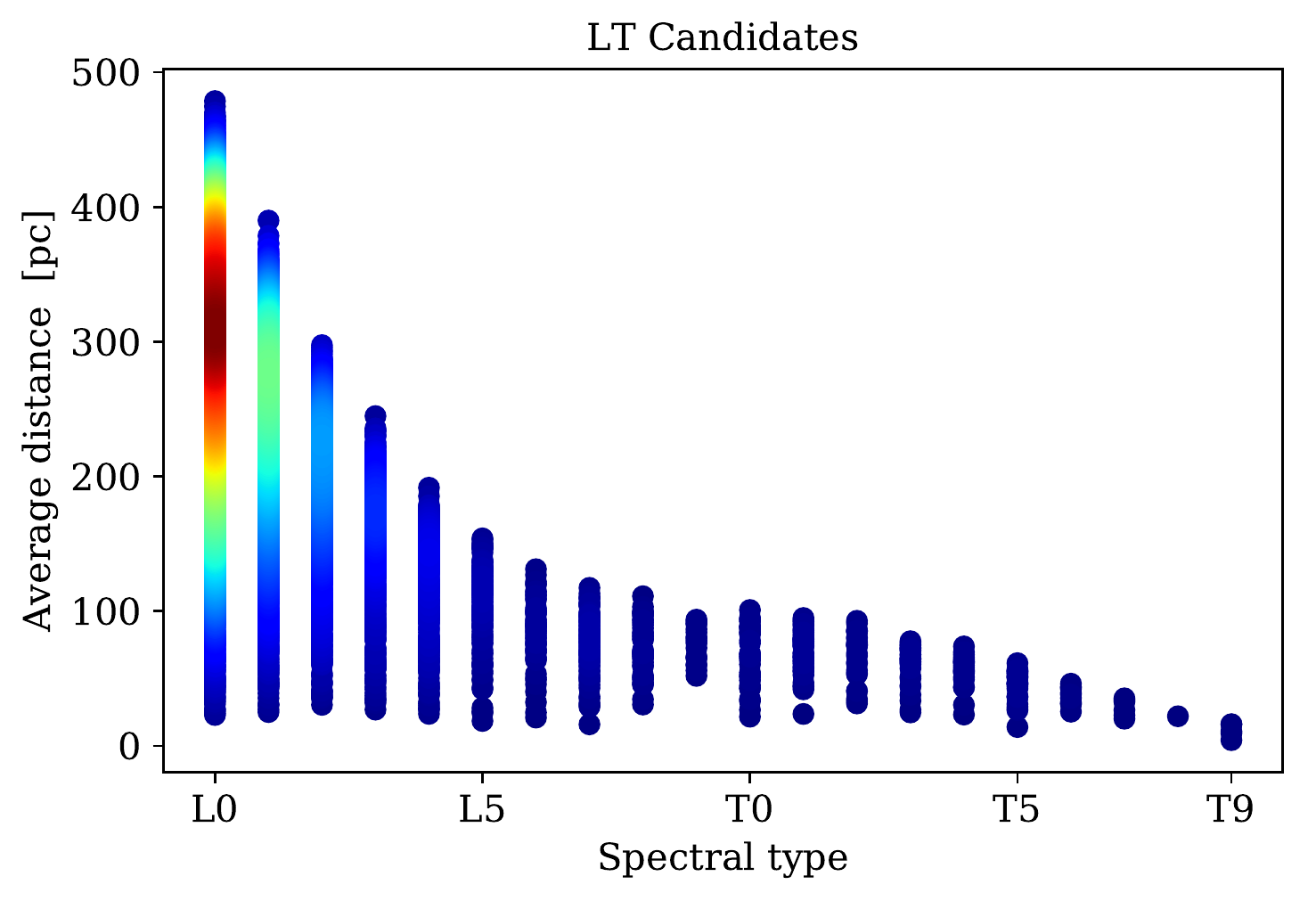}
\caption{Distances as a function of spectral type. Distances have been calculated using the average value from the distance modulus obtained using all available bands. The colour scale represents the density. Most LT candidates are early L types at distances smaller than 500 parsecs.}
\label{fig:distancecandidates}

\end{figure}

\begin{figure}
    \includegraphics[width=\linewidth]{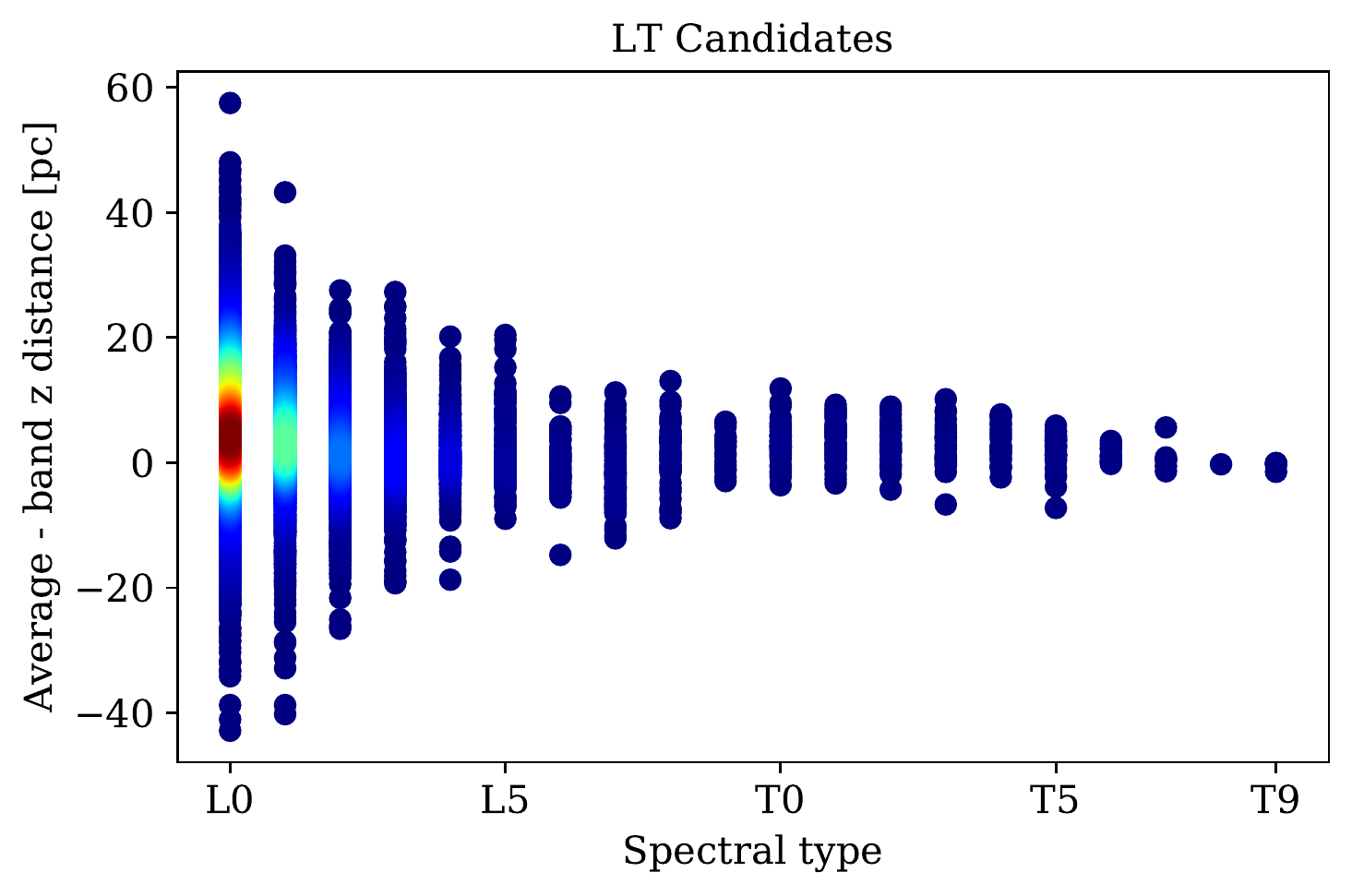}
\caption{Difference between the two distance estimators. In one where we only use $z$ band and another where the average value over all available bands is used. The colour scale represents the density.}
\label{fig:distancestd}

\end{figure}

Absolute magnitudes have been anchored to \textit{M\_W1} and \textit{M\_W2} from \citet{Dup12} as explained in Section~\ref{sec:code}. We compare two estimates for the distance: one where we use all available bands from the $i,z,Y,J,H,K_{s},W1,W2$ set, and then we average over the bands to give a mean value, and the other where we use band $z$ only. The distance distribution can be found in Fig.~\ref{fig:distancecandidates} for the averaged value. In Fig.~\ref{fig:distancestd} we show the difference between the averaged value and the $z$ band estimate. In general, both definitions agree. In the published catalogue both estimates are given with names \verb+DISTANCE_AVG+ and \verb+DISTANCE_Z+.

\section{Electronic Catalogue}
\label{app:catexplanation}

This is the largest photometric LT catalogue published to date, containing 11,745 sources. We also publish the M dwarf catalog in a separate table containing 20,863 sources. Table~\ref{tab:mltcatalog} shows a sample of the LT catalogue, which can be accessed at \url{https://des.ncsa.illinois.edu/releases/other/y3-mlt}. The number of LT types found in the target sample is subject to completeness and purity effects (see Section~\ref{sec:systematicsgalmod}). 

Spectral type is given by \verb+SPT_PHOT+ with the following convention: L types are assigned \verb+SPT_PHOT+ from 10 to 19, corresponding to L0 to L9, and from 20 to 29 for the T types, corresponding with T0 to T9. In the M dwarf catalogue, M types are assigned \verb+SPT_PHOT+ from 6 to 9. We also give the $\chi^{2}$ of the classification and the number of bands used with columns \verb+XI2_CLASSIF+ and \verb+NBANDS+. Distances are provided with 2 estimates, one based on band $z$ only (\verb+DISTANCE_Z+) and another based on the average of the distances calculated in all the available bands (\verb+DISTANCE_AVG+). We also tag if the source is peculiar based on a z-score analysis of its $\chi^{2}$, where \verb+PECULIAR+=1 is peculiar and \verb+PECULIAR+=0 is not. Distances and the peculiar tag are only present in the LT catalogue. The LT and the M Catalogues also include $i,z,Y$ magnitudes from DES, $J,H,K_{s}$ from VHS and $W1,W2$ from AllWISE as used in the classification. DES magnitudes are corrected with updated zeropoints different from those present in the public DR1 release \citep{abb18} and they will be published in the future.

We matched the M and LT catalogues with Gaia DR2 release~\citep{gaia}, but the amount of matches was very low in the LT catalogue. Only 282 matches were found (2.4\%). In the M catalogue, we found 1,141 matches (5.4\%), but these spectral types are not the scope of this paper.

\begin{landscape}
\thispagestyle{empty}

\begin{table}
\caption[]{Example of the LT catalogue published, distances and peculiar flag are only present in the LT catalogue. SPT\_PHOT is the spectral type, NBANDS the number of bands used in classification and XI2\_CLASSIF the $\chi^{2}$ for the best-fit in \textit{classif}. DISTANCE\_AVG is the average distance calculated from apparent magnitudes in all the available bands (in parsecs) and DISTANCE\_Z the distance calculated from the apparent magnitude in DES z band (in parsecs). HPIX\_512 is the \textit{Healpix} pixel ID for $order=Nest$ and $nside=512$ and PECULIAR is a category given if it has a $\chi^{2} > 3.5\sigma$ median absolute deviations (1 = peculiar,  0 = no peculiar).}
\label{tab:mltcatalog}
\begin{tabular}{llllllllll}
\hline
 COADD\_OBJECT\_ID &          RA &        DEC &  SPT\_PHOT &  NBANDS &  XI2\_CLASSIF & DISTANCE\_AVG   &  DISTANCE\_Z &    HPIX\_512 & PECULIAR \\
  
\hline
293254901 &   16.568 & -52.537 &       8 &     5 &     4.326 &    419.848 &  406.993 &  2240708 & 0 \\
302513036 &   18.383 & -51.732 &      10 &     5 &     5.944 &    277.501 &  270.822 &  2240815 & 0 \\
232352550 &   23.216 & -63.690 &       8 &     5 &     3.974 &    380.790 &  370.102 &  2142795 & 0 \\
237139089 &  313.501 & -60.505 &       8 &     5 &     1.837 &    582.501 &  555.256 &  2935140 & 0 \\
204133552 &  322.767 & -53.430 &       9 &     5 &     4.503 &    491.583 &  495.687 &  2939932 & 0 \\
 71547410 &  330.937 & -57.231 &      10 &     7 &     3.993 &    308.368 &  306.911 &  2915933 & 0 \\
185491825 &  315.506 & -53.252 &       9 &     5 &     1.530 &    393.643 &  398.255 &  2945055 & 0 \\
 93231446 &   33.147 & -51.827 &       8 &     7 &     0.971 &    262.541 &  259.220 &  2244895 & 0 \\

\hline
\end{tabular}
    \end{table}

    \begin{table}
\contcaption{DES magnitudes are in AB. $i,z$ are PSF magnitudes from ``SOF" while $Y$ band is from ``Sextractor". VHS magnitudes are in Vega and are aperture magnitudes at 2 arcseconds.}
    \label{tab:mltcatalogcontinue1}
\begin{tabular}{llllllll}
\hline
                   PSF\_MAG\_I &  PSF\_MAGERR\_I &  PSF\_MAG\_Z &  PSF\_MAGERR\_Z &  AUTO\_MAG\_Y &  AUTO\_MAGERR\_Y &  JAPERMAG3 &  JAPERMAG3ERR  \\
     
\hline
21.971 &      0.029 &  20.735 &      0.016 &   20.580 &       0.088 &  18.937 &      0.039 \\
22.571 &      0.041 &  20.937 &      0.022 &   20.394 &       0.068 &  18.722 &      0.037 \\
21.738 &      0.028 &  20.529 &      0.014 &   20.292 &       0.088 &  18.588 &      0.050 \\
22.800 &      0.055 &  21.410 &      0.040 &   21.156 &       0.159 &  19.466 &      0.086 \\
22.893 &      0.051 &  21.603 &      0.031 &   21.389 &       0.172 &  19.693 &      0.108 \\
22.736 &      0.043 &  21.209 &      0.027 &   20.882 &       0.088 &  18.803 &      0.055 \\
22.488 &      0.033 &  21.128 &      0.018 &   20.768 &       0.089 &  18.949 &      0.055 \\
21.069 &      0.013 &  19.756 &      0.007 &   19.467 &       0.029 &  17.857 &      0.031 \\

\hline
\end{tabular}
    \end{table}

  \begin{table}
\contcaption{VHS magnitudes are in Vega and are aperture magnitudes at 2 arcseconds. WISE magnitudes are in Vega and are magnitudes measured with profile-fitting photometry. For more details visit \url{https://des.ncsa.illinois.edu/releases/other/y3-mlt}.}
    \label{tab:mltcatalogcontinue2}
\begin{tabular}{llllllll}
\hline
 HAPERMAG3 &      HAPERMAG3ERR &   KSAPERMAG3 &  KSAPERMAG3ERR &       W1MPRO &  W1SIGMPRO &       W2MPRO &  W2SIGMPRO   \\
\hline
 
 -9999. &  -9999. &   17.921 &       0.105 & -9999.&  -9999. & -9999. &  -9999. \\
-9999. &  -9999. &   17.773 &       0.108 & -9999. &  -9999. & -9999. &  -9999.\\
-9999. &  -9999. &   17.937 &       0.156 & -9999. &  -9999. & -9999. &  -9999.\\
-9999. &  -9999. &   18.868 &       0.329 & -9999. &  -9999. & -9999. &  -9999.\\
-9999. &  -9999. &   18.266 &       0.178 & -9999. &  -9999. & -9999. &  -9999.\\
   18.309 &      0.078 &   17.860 &       0.106 &    17.604 &      0.175 &    17.292 &  -9999.\\
-9999. &  -9999. &   18.098 &       0.157 & -9999. &  -9999. & -9999. &  -9999.\\
-9999. &  -9999. &   16.835 &       0.060 &    16.702 &      0.071 &    16.476 &      0.189\\

\hline
\end{tabular}
    \end{table}

\end{landscape}

\section{Modeling the number counts of LT dwarfs}
\label{sec:simulation}

This section describes the effort to create robust expected number counts of LT dwarfs. The algorithm is called \textit{GalmodBD}. It is a \verb+Python+ code that computes expected galactic counts of LT dwarfs, both as a function of magnitude, colour and direction on the sky, using the fundamental equation of stellar statistics. It was adapted from the code used by \citet{San96} and \citet{Ker01} to model HST star number counts, and by \citet{San10} for a preliminary forecast of DES star counts. In the current analysis, we kept the density laws for the different Galactic components, and simply replaced the specific luminosity functions of normal stars by the BD number densities as a function of the spectral type presented in Subsection~\ref{sec:code}. 

\textit{GalmodBD} also creates synthetic samples of LT dwarfs based on the expected number counts for a given footprint. Besides a model for the spatial distribution of BDs, \textit{GalmodBD} uses empirically determined space densities of objects, plus absolute magnitudes and colours as a function of spectral type. The model is described in Subsection~\ref{sec:code}. The point generating process is described in Subsection~\ref{sec:synth}. The validation tests are provided in Appendix~\ref{app:codeval}.

\subsection{\textit{GalmodBD}}
\label{sec:code}

For convenience, we will refer to the space density versus spectral type relation as the luminosity function (LF), which is somewhat of a misnomer, since luminosity does not scale uniquely with spectral type in the LT regime. We refer to the colours versus spectral type relations as C-T relation. The code requires several choices of structural parameters for the Galaxy, such as the density law and local normalisation of each Galactic component. Equally crucial are parameters that govern the region of the sky and magnitude and colour ranges where the expected counts will be computed.

These parameters are listed in different configuration files. Currently, only one choice of LF is available, taken from Table 8 of \citet{Mar15}. More specifically, for types earlier than L4, we use \citet{Cru07} space density values; from L4 to T5 we use those of \citet{Mar15} themselves, and for later types than T5 we use \citet{Bur13}. 

For the C-T relations, we first build absolute magnitude vs. spectral type relations for AllWISE data. \textit{M\_W1}, and \textit{M\_W2} versus type relations are taken from Figures 25, 26 and Table 14 of \citet{Dup12}. Once we anchor absolute magnitudes in these bands, we use the C-T relations found in Section~\ref{sec:calib} in Table~\ref{tab:galmodcolors}, Fig.~\ref{fig:sptcolor} to populate \textit{M\_i,M\_z,M\_Y,M\_J,M\_H,M\_Ks}.

The expected number counts are computed by direct application of the fundamental equation of stellar statistics. In summary, given a choice of apparent magnitude range in some filter, and some direction in the sky (pointer), we go through distance steps and for each of them, find the range of LT spectral types whose absolute magnitudes fit into the chosen apparent magnitude range. We then compute the volume element in the selected direction and distance and multiply it by the appropriate LF value.

The final number count as a function of apparent magnitude is the sum of these contributions for all appropriate combinations of spectral types (through their associated absolute magnitudes) and distances. Because we also have colour versus type relations, we can also perform the same integral over distance and type range to compute number counts as a function of colour.

\textit{GalmodBD} incorporates extinction and dereddened using the \citet[SFD98]{Sch98} dust maps, although it can also use~\citep{Bur82} maps. Conversion from A\_V and E(B-V) to SDSS A\_i, A\_z and E(i-z) are based on \citet{Ann08}. Nonetheless, in this first application of the code, we have not included any reddening effect since our sample is concentrated in the solar neighborhood. 

The code currently permits many choices of magnitudes and colours for number counts modeling, including colours in SDSS and DES in the optical, VHS and 2MASS in the near-infrared and AllWISE in the infrared. 

In Appendix~\ref{app:codeval} we validate the \textit{GalmodBD} simulation comparing its predictions with a prediction based on a single disk component analytical count. 

\section{Synthetic catalogues}
\label{sec:synth}

Besides determining expected N(m) and N(col) for BDs within some magnitude range and over some chosen area on the sky, the code also generates synthetic samples based on these number counts. This is done for every chosen direction, distance and type (absolute magnitude) by randomly assigning an absolute magnitude within the range allowed by the spectral type bin, and then converting it to apparent magnitude. We use the same random variable to assign absolute magnitudes within each bin for all filters, in order to keep the synthetic objects along the stellar locus in colour-colour space. 

For a complete description, we also need to assign random magnitude errors to mimic an observed sample, which will depend on the particular case. Therefore, error curves as a function of magnitude for each filter are required. In subsection~\ref{sec:mimicdes} we detail this step for our sample. By default, \textit{GalmodBD} accepts an exponential error distribution. In case of more sophisticated models, the error must be introduced afterward.

For synthetic dwarfs, the code outputs its Galactic component, spectral type, Galactic coordinates (\textit{l,b}) of the centre of the pointer, distance and  magnitudes in the set of filters, for example: $i_{DES}$, $i_{SDSS}$, $z_{DES}$, $z_{SDSS}$, $Y_{DES}$, $J_{VHS}$, $J_{2MASS}$, $H_{VHS}$, $H_{2MASS}$, $K_{s,VHS}$, $K_{s,2MASS}$, $W1$ and $W2$. Both the true and observed magnitudes, as well as their errors, are output, regardless of the chosen pair (mag, colour) used in the output expected model counts. This latter choice, coupled with the apparent magnitude range and the direction chosen, however, affects the total number of points generated.

In order to have a synthetic sample with coordinates, we randomly assign positions to the sources within the given pointer area (see Section \ref{sec:footpointers}).

These simple synthetic catalogues can be compared to a real catalogue of BD candidates by correcting the expected numbers for the visibility mask, that accounts for catalogue depth and detection variations, and for estimates of purity and completeness levels of the observed catalogue.

\section{\textit{GalmodBD} in the target sample}
\label{sec:galmodbdindes}

In this section we detail the input information used to feed the \textit{GalmodBD} simulation that reproduces the $DES \cap VHS \cap AllWISE$ data. Besides the empirical colour template relations, we need to define the footprint of the simulation and the photometric error model to produce realistic MLT catalogues. 

\subsection{Footprint and pointers definition}
\label{sec:footpointers}

To create a sample that resembles the target sample, we create a grid of pointers following the DES tile distribution.

DES coadd data are divided into square (in spherical coordinates) regions of equal area, covering $0.534 \ deg^{2}$ each, called tiles. We use this information to define our pointers covering the area occupied in Fig.~\ref{fig:y3footprint}. The pointers defined to run \textit{GalmodBD} have the same coordinates as the centre of the DES tiles and the same area of $0.534 \ deg^{2}$. Later, tiles are intersected with the VHS footprint in the DES area, covering the brown areas in the same figure. Eventually, we end with 5187 pointers of $0.534 \ deg^{2}$, covering $2,374 \ deg^{2}$.

After running the simulation in each pointer, we assign random positions within the area of the pointer to each object in the simulated catalogue. At this point we would need to consider the effect of incompleteness in the footprint, i.e., apply the footprint mask. 

To study the effect the mask might have in the number of BDs recovered by our method, for each \textit{GalmodBD} run, we create multiple synthetic position catalogues and pass the data through the footprint mask. Eventually, obtaining a statistic of the mean and standard deviation (std) of the number of MLT sources that we will lose. In our case we run 500 realizations for each \textit{GalmodBD} model to calculate the effect on footprint completeness in Section~\ref{sec:systematicsgalmod}.

\subsection{Mimic DES photometric properties}
\label{sec:mimicdes}

Another ingredient in the \textit{GalmodBD} simulation is the photometric errors that apply to the simulated data to create an observed synthetic sample.

To model the signal-to-noise distribution we start by selecting a random sample of the $DES \cap VHS \cap AllWISE$ data, limited to $z<23$ and selecting point sources only using the extended classification from DES Y3 data. We next apply the following algorithm for each band: 

First, we divide the sample in magnitude bins of width=0.1. For each bin, we estimate the probability density function of the magnitude error using a kernel density estimation (\verb+KDE+). 

With the \verb+KDE+ information for each thin magnitude bin, we can assign a magnitude error for a given magnitude with a dispersion that follows the \verb+KDE+. Once an error is assigned to a source, we estimate apparent magnitudes assuming a Gaussian distribution centred in the true apparent magnitude and with a standard deviation equals to their magnitude error. 

We extract the \verb+KDE+ for magnitude bins where we have more than 60 objects. In the extreme cases of very bright or very faint objects in the magnitude distribution, this requirement is not met. Therefore we expand the distribution along the bright end by repeating the \verb+KDE+ from the brightest bin with enough statistics. In the faint end, we fit the mean and sigma of the faintest 4 bins with enough statistics by a second-order polynomial and extrapolate this fit towards fainter magnitudes. 

\begin{figure*}
\begin{center}
    \includegraphics[width=0.99\linewidth]{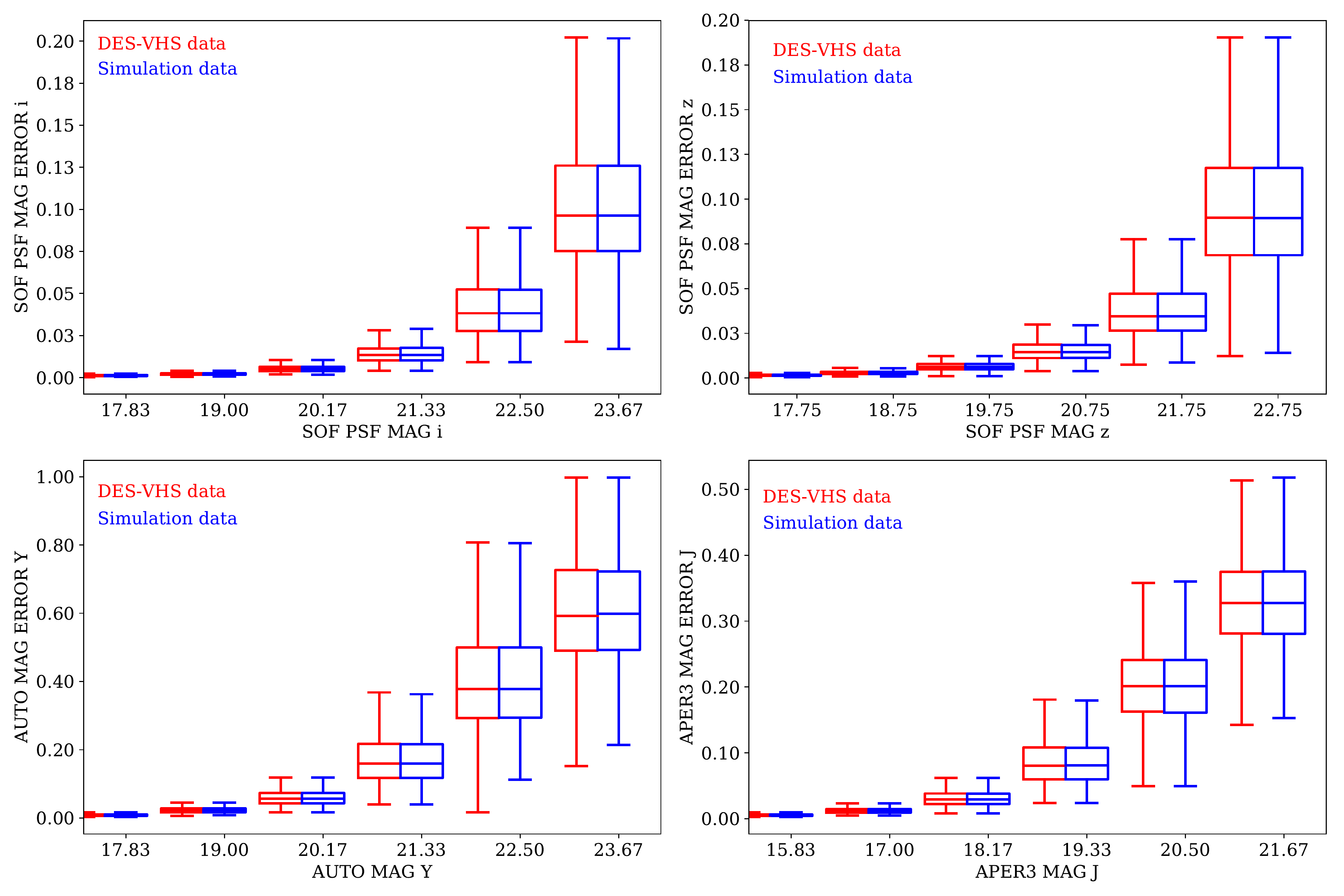}
\end{center}
\caption{Box-plot comparing the magnitude uncertainties as a function of magnitude for DES $i,z,Y$ and VHS $J$ data, versus the modeled distribution in \textit{GalmodBD} simulation. In red, the $DES \cap VHS$ distribution, in blue, the simulated distribution. The horizontal line within the box indicates the median, boundaries of the box indicate the 25th- and 75th-percentile, and the whiskers indicate the highest and lowest values of the distribution. The median and dispersion agree very well for all bands.}
\label{fig:errorcurves1}
\end{figure*}

In Fig.~\ref{fig:errorcurves1} we summarize the error modeling for the bands of interest: $i,z,Y$ and $J$. In the Figures, we compare the error distribution as a function of magnitude for the real data, individually for each band, with the simulated data. The median and dispersion agree very well for all bands.

\section{\textit{GalmodBD} results}
\label{sec:galmobdresults}

In Table~\ref{tab:galmodgrid} we show the result of running \textit{GalmodBD} for different galactic models varying the thin disk scale height at $h_{z,thin}=250, 300, 350, 400, 450, 500, 550 \ pc$ and the local thick-to-thin disk density normalisation $n_{sun,thick}= 5, 10, 15, 20 \%$ defined as $n_{sun,thick}=\rho_{thick}(R_{\sun})/\rho_{thin}(R_{\sun})$. After some testing, it was clear the LT number counts were most sensitive to these two parameters. We ran the simulation for M dwarfs ranging M0 to M9 and BDs from L0 to T9 (even though the scale height of LT types does not need to be the same as for M types). 

We run \textit{GalmodBD} up to a true apparent magnitude limit of $z_{true}<25$. Next, we assign errors and observed magnitudes as explained in Section~\ref{sec:mimicdes}. Later, we select MLT with $z_{obs}<=22$ and $(SNR) > 5\sigma$ in $z,Y,J$ to build our simulated MLT parent catalogues. Using these catalogues, we can study the completeness and purity of our sample (see Section~\ref{sec:systematicsgalmod}). 

In Table~\ref{tab:galmodgrid}, the third column is the number of M types for a given model that pass the colour cuts defined in Section~\ref{sec:targetselection} and $z_{obs} < 22$ and $(SNR) > 5\sigma$ in $z,Y,J$. The fourth column is the number of LT types when $z_{obs} < 22$ and $(SNR) > 5\sigma$ in $z,Y,J$. We have not applied here the colour cut, since it is a source of incompleteness and we treat it together with other effects in Section~\ref{sec:systematicsgalmod}. 


\begin{table*}
\centering
\caption{Number of BDs as a function of MW structural parameters in \textit{GalmodBD}. $h_{z,thin}$ is the thin disk scale height in parsecs, $n_{sun,thick}$ is the local thick-to-thin disk density normalisation in percentage. The third column is the number of M types after the magnitude limit cut, the signal-to-noise cut and the colour cut. The fourth column is the number of LT types applying only the magnitude limit cut and the signal-to-noise cut and the fifth column is the number of early L types defined as those with spectral types less than L4. This last column will be compared with the number of BDs detected in our sample to give a first estimate of the thin disk scale height for early L types.}

\label{tab:galmodgrid}
\begin{tabular}{| c  c | c cc  }
        \hline \hline
    
$h_{z,thin}$ &  $n_{sun,thick}$ &  M types  & LT types & Early L types  \\
 pc &  \% & [$snr>5\sigma$ + $z<22$ + colour cut] & [$snr>5\sigma$ + $z<22$] & (L0, L1, L2, L3)  \\
 \hline
 
250 & 5  & 11,609  &8,403  &  7,427\\ 
250 & 10  &  12,869  &9,094 &  8,055\\ 
250 & 15  & 14,121  &9,742   & 8,663\\ 
250 & 20  & 15,428  &10,475  & 9,333\\ 
\hline
300 & 5  & 13,782  & 9,426  &8,381\\ 
300 & 10  & 15,113  &10,086 &  8,992\\
300 & 15  & 16,366  &10,821 &  9,663\\ 
300 & 20  & 17,652  &11,427 &  10,211\\ 
\hline
350 & 5  &  15,768  &10,197 & 9,118\\
350 & 10  & 17,018  &10,887  & 9,758\\
350 & 15  &  18,173  &11,507 & 10,336\\ 
350 & 20  & 19,599  &12,151  & 10,929\\ 
\hline
400 & 5  & 17,329 &10,891   &9,798\\ 
400 & 10  & 18,611  &11,590 &  10,454\\ 
400 & 15  & 19,968  & 12,313 &  11,079\\ 
400 & 20  & 21,345  &12,958 &  11,684\\ 
\hline
450 & 5  & 18,831  & 11,463  & 10,343\\ 
450 & 10  & 20,015  & 12,022 &  10,873\\ 
450 & 15  & 21,486  &12,774 &  11,575\\ 
450 & 20  & 22,653  &13,384 &  12,115\\ 
\hline
500 & 5  &  20,231  &11,891  & 10,764\\
500 & 10  & 21,425  & 12,589 & 11,409\\ 
500 & 15  & 22,732  &13,255  & 12,028\\
500 & 20  & 24,086  &14,100  &12,785 \\ 
\hline
550 & 5  &  21,217  &12,297  & 11,152\\
550 & 10  & 22,551  & 13,038 & 11,811\\ 
550 & 15  & 23,859  &13,763  & 12,501\\
550 & 20  & 25,121  &14,359  & 13,033\\ 
\end{tabular}
\end{table*}

\section{Purity and completeness}
\label{sec:systematicsgalmod}

In this section we detail the various sources of error in the measurement of the number counts of LT dwarfs from the target sample. We will use a combination of both the calibration samples and the \textit{GalmodBD} simulation. There are two issues to consider: one is to identify and quantify contamination and incompleteness effects on our sample, so we can correct the simulated numbers in order to compare them to the data. The other is assessing the uncertainty associated with these corrections and the expected fluctuations in the corrected number counts. The effects we consider are:
\renewcommand{\labelitemi}{\textendash}
\begin{itemize}
\item Colour-colour incompleteness: how many LT we lose by applying the colour cut to select candidates.
\item Loss of targets from the catalogue due to proper motions.
\item Footprint effects: how many LT we lose due to masking effect.
\item Loss of LTs due to misclassification.
\item Contamination of targets due to unresolved binary systems in our magnitude-limited catalogue.
\item Contamination by M dwarfs and extragalactic sources.
\end{itemize}


\subsection{Incompleteness}

To assess how many BDs we miss due to the colour-colour selection, we look at the number of LT types that do not enter our colour selection both in the Gagn\'e sample as well as in the \textit{GalmodBD} simulation. The colour range selected was chosen to minimize incompleteness. But peculiar early L dwarfs may eventually be found outside our colour range. In the Gagn\'e sample, from the list of 104 known BDs (including young types), 2 are left outside the colour range (2\%). In \textit{GalmodBD}, we haven't modeled peculiar BDs and therefore, the incompleteness level should be low. Applying the colour cut to the \textit{GalmodBD} outputs, as expected, led to a mean completeness of 98.8\%. We define the colour-colour completeness correction to be the average of the two. The uncertainty associated with this correction should be low. Conservatively, we define the uncertainty to be of 1\%, leading to $C_{colour} = 98.4 \pm 1.\%$.

In order to estimate the number of missing sources due to proper motions, we will assume that the mean proper motion of LT types decreases with distance. Considering a conservative 3-year difference in astrometry between DES and VHS, a $2\arcsec$ matching radius should be complete for distances $> 50 \ pc$. In fact, looking at the BDs that we recover visually beyond the $2\arcsec$ radius in Section~\ref{subsec:gagne} and that have distances in the Gagn\'e sample, more than 95\% of the missing BDs have a distance $<50 \ pc$. So we can set this as an upper limit for this effect. Above $50 \ pc$, we match at $2\arcsec$ almost all of the BDs presented in the calibration sample. Likewise, we do match some of the BDs below $50 \ pc$, at least $20\%$ of them. 

In summary, matching the Gagn\'e sample to DES within $2\arcsec$ we recover $20\%$ of the BDs below $50 \ pc$ and $100\%$ above that distance. This is a conservative limit, but it gives us a sense of the percentage loose by proper motions. Note that we have not considered here that we might miss some of the targets due to the incompleteness of the footprint so the 20\% should be higher.

If we now look at the \textit{GalmodBD} simulation, we predict the number of LT types with distances $< 50 \ pc$ to be $2\%$ of the whole sample. Considering an $80\%$ completeness and an error in the determination comparable to the effect itself, it results in a proper motion completeness correction of $C_{pm}=98.4 \pm 2\%$.

In terms of footprint incompleteness, we apply the algorithm presented in Section~\ref{sec:footpointers} to estimate the effect of the footprint mask: we produce 500 realizations of each \textit{GalmodBD} model and estimate the mean and std of LT type number counts that survive the masking process. We then average over all models to obtain a model-independent completeness correction of $C_{foot} = 93.6\pm 3.5\%$.

Finally, there is the effect of misclassification of LT types as M types. We estimated the incompleteness of LT types due to misclassification as $C_{classif} = 85 \pm 1\%$. In the next subsection, we describe this correction along with the corresponding contamination effect, namely the misclassification of M dwarfs as LT types.

\subsection{Contamination}
\label{subsec:contamina}

Our LT dwarfs catalogue is limited to $z \leq 22$. Unresolved binary systems containing two LT dwarfs will have a higher flux than a single object and hence will make into the catalogue even though each member of the system individually would not. This boosts the total number of LTs in our sample. We estimate the effect of unresolved binarism systems as follows. According to \citet{luh12} (and references therein), the fraction of binaries decreases with the mass of the primary, while the mass ratio ($q=m1/m2$) tends to unity (equal-mass binaries). In our case, we are only interested in those binaries where the primary is an LT. Binary systems with the LT as secundary, will have a small mass ratio ($q<<1$), and therefore, they will be uncommon. Also, we cannot census this type of population since the light from the primary will dominate the light from secondary. \citet{luh12} quotes different estimates the fraction of binaries where the primary is an LT, $f_{bin}$. Observational data suggest $f_{bin} \simeq 7\%$, but may be prone to incompleteness, especially for close pairs. Theoretical models of brown dwarf formation predict $f_{bin}$ in the $20\% - 30\%$ range \citep{Maxted2005,Basri2006}. Nonetheless, a recent estimate by \citet{font18} would confirm the empirical trend of \citet{luh12} and therefore we here adopt $f_{bin}=0.07$. If we assume that $q\sim 1$ in all cases, we will obtain binary systems that have the same spectral type as the primary, but with a magnitude that is 0.75 times brighter (since the system flux is twice that of the primary). To estimate the contamination by binaries in the target sample, we follow the description of ~\citet{bur10}. According to equation 3 in that paper, the correction factor is:

\begin{equation}
P_{binary} = \frac{\gamma -1}{\gamma + 1/f_{bin} - 1}
\label{eq:fbin}
\end{equation}

where $\gamma= 2\sqrt{2}$ for equal mass/luminosity binaries, and $f_{bin} = 0.07$. Finally, we get $P_{binary} = 111 \pm 1\%$.

There are two additional sources of contaminants applied here: one is the migration of M dwarfs that have been wrongly assigned as LT type after running \textit{classif}, and the other is the contamination by extragalactic sources.

Ideally, in order to estimate the contamination by M dwarfs and the incompleteness of LT types due to the photometric classification, we should run \textit{classif} on the output of \textit{GalmodBD}. Unfortunately, our classification code is fed with the same templates used in the simulation and, therefore, the classification uncertainty we get by running \textit{classif} in \textit{GalmodBD} is unrealistically low. In order to estimate the number of M dwarfs that are classified as LT types and vice versa, we perturb the true spectral types for \textit{GalmodBD} sources following a normal distribution with dispersion given in Section~\ref{sec:classifcalib}: $\sigma_{M}=0.69$, $\sigma_{L}=1.03$ and $\sigma_{T}=1.12$ and mean value centred in the true spectral type. 

Once we perturb the true spectral types to get a pseudo-photometric calibration in \textit{GalmodBD}, we can estimate how many LTs will be classified as M, and vice versa. We estimate this number for all the \textit{GalmodBD} realizations and find that $\sim 11 \%$ of M dwarfs (after the colour cut) are given LT type. This corresponds to contamination by about 14-19\% of the corresponding LT simulated sample. This means a mean LT purity level of $P_{class}=83\pm 2\%$. We also use the same procedure applied on the \textit{GalmodBD} data to assess the number of LT types migrating to M type. This is an additional incompleteness of $C_{class} = 85 \pm 1\%$, to be added to those from the previous subsection.

The removal of extragalactic contaminants is another source of uncertainty. We saw in Section~\ref{mainclassif} that $\lesssim 10\%$ of the MLT sample have galaxy or quasar class. These are not modeled in \textit{GalmodBD}, and therefore we do not need to apply a correction due to extragalactic contamination, although there will always be uncertainty associated with it. Based on tests with different \verb+Lephare+ SED libraries and the ambiguity in classification for sources with \verb+NBANDS+ < 5, we estimate a standard deviation associated with the removal of extragalactic sources of $\sigma_{SG} \sim 5.\%$.

\subsection{Completeness and purity summary}

Summarizing all these effects, we estimate the completeness and contamination up to $z<22$ at $5\sigma$ as:

\begin{itemize}

\item Completeness due to the colour-colour cut of $C_{colour} = 98.4 \pm 1.\%$.

\item Completeness due to proper motions of $C_{pm} = 98.4 \pm 2\%$.

\item Completeness due to masking of $C_{foot} = 93.6\pm 3.5\%$.

\item Completeness due to LT misclassified as M types of $C_{class} = 85 \pm 1\%$.

\item Contamination due to unresolved binary systems measured as the fraction $f_{bin} = 7\%$ of sources: $K_{bin}=11\pm 1\%$. 

\item Contamination of the LT sample due to M types classified as LTs: $K_{class}=17\pm 2\%$.

\item Uncertainty in the extragalactic contamination removal algorithm introduces an additional systematic error of $\sigma_{extra} = 5.\%$.

\end{itemize}

Combining all these effects, we define a correction factor:

\begin{equation}
    CP_{LT} = \frac{C_{colour}*C_{pm}*C_{foot}*C_{class}}{(1-K_{bin})*(1-K_{class})}
    \label{eq:CPLT}
\end{equation}

that multiplies the LT sample. This correction factor accounts for the effect of incompleteness and purity, and applies to our definition with $z<22$ and $5\sigma$ in $z,Y,J$. Different magnitude limits would lead to different corrections. In this case, the final correction factor is $C_{LT} = 1.05$. This factor is applied to the number of detected LT $N(LT)_{obs} = 11,745$ to get an estimate of the total number of LT in the footprint up to $z<22$ with at least a $5\sigma$ detection in $z,Y,J$ as $N(LT)_{true}=N(LT)_{obs}*C_{LT}$. At first approximation, the uncertainty in the number of LT types is the square root of the sum of the squares of the individual errors listed above:

\begin{multline}
\sigma_{LT}= \\
\sqrt{\sigma_{col}^2+\sigma_{pm}^2+\sigma_{bin}^2+\sigma_{foot}^2+\sigma_{cl,C}^2+ \sigma_{cl,P}^2 + \sigma_{ex}^2} 
\label{eq:sigmamlt}
\end{multline}

with $\sigma_{LT} \sim 7\%$. Finally, the number of LT's to compare against the different SFH scenarios is in round numbers $N(LT) \sim 12,300 \pm 900$. If we only select the early L types (L0, L1, L2 and L3) for which we want to estimate the thin disk scale height, the number to compare with becomes $N(L_{0,1,2,3}) \sim 11,500 \pm 800$.

\section{The thin disk scale height}
\label{sec:sim_des}

\citet{jur08} estimate the thin disk scale height and the local thick-to-thin disk density normalisation of SDSS M dwarfs up to a distance of $2 \ kpc$. They found $h_{z,thin} = 300 \ pc$ and $n_{sun,thick} = 12\%$.

As a first application of the brown dwarf catalogue and \textit{GalmodBD}, we compare the number of observed LT types with different realizations of \textit{GalmodBD} to shed light on the thin disk scale height of the LT population. Since T types are less than 2\% of the sample and go only up to $100 \ pc$, in practice, we are estimating the scale height of the early L types. A previous attempt to measure the thin disk scale height of L types can be found in \citet{rya05}, where they estimate a value of $h_{z,thin}=350 \pm 50$ using only 28 LT types and on a very simplistic Galactic model. Since our model is more detailed and we have a much higher statistic, we believe that our results will be more reliable. More recently, \citet{Sor18} have used data from the first data release from the Hyper Suprime Camera \citep[HSC, ][]{Miy18,Aih18} to estimate the vertical thin disk scale height of early L types with an estimate between 320 pc and 520 pc at 90\% confidence.

\begin{figure}
\includegraphics[width=\linewidth]{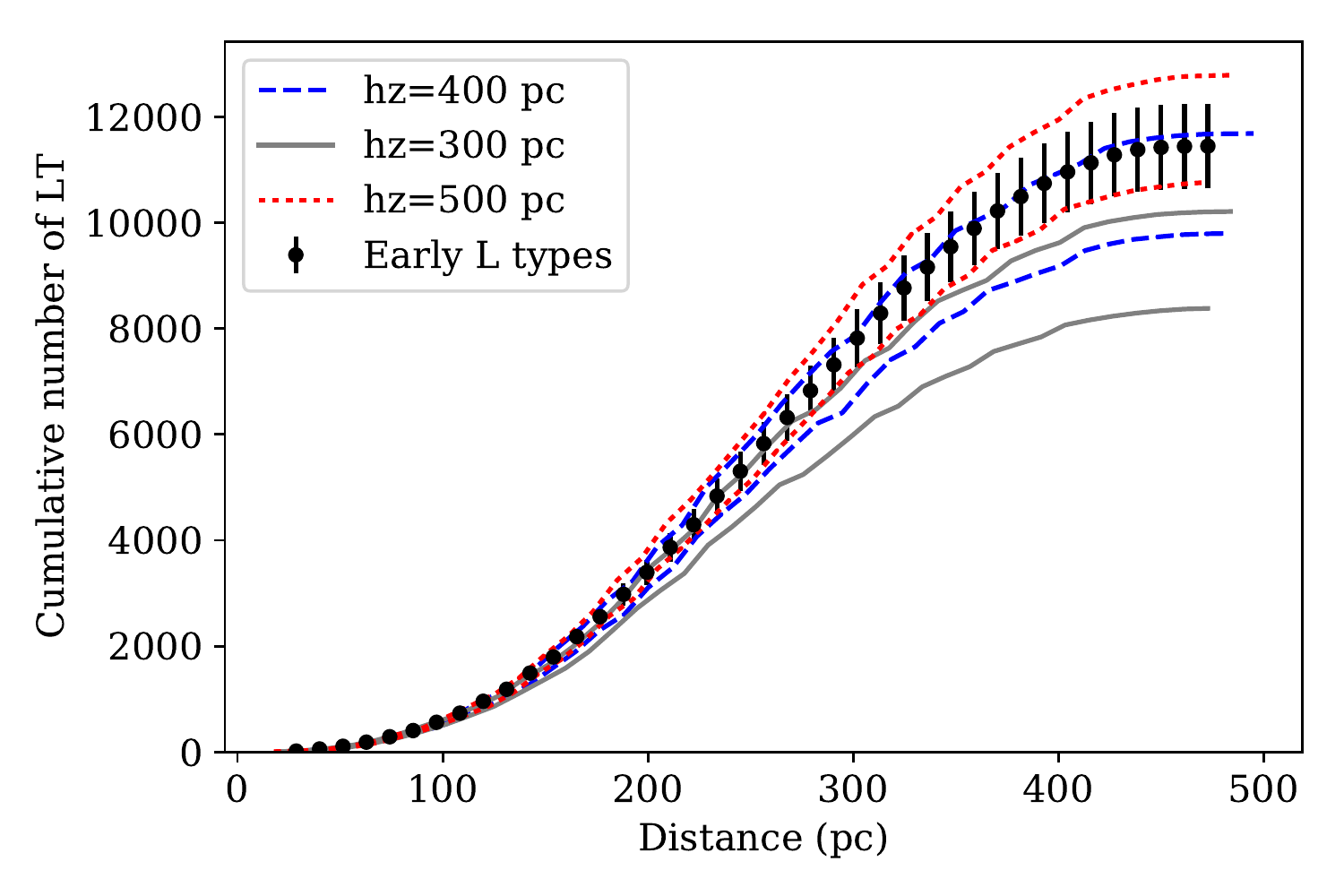}
\caption{Comparison of the number counts of early L types (L0, L1, L2 and L3) with three models of the thin disk scale height. In Grey we show models with a scale height similar to that for M dwarfs \citep{jur08} with $h_{z}=300 \ pc$, in dashed blue, models with $h_{z}=400 \ pc$ and in red, models with $h_{z}=500 \ pc$. For each model we show two estimates, one where $n_{sun,thick}=5\%$ (lower limit) and another where $n_{sun,thick}=20\%$ (upper limit).}
\label{fig:scaleheight}
\end{figure}

Comparing the value of $N(LT) \sim 12000 \pm 900$ with the grid of \textit{GalmodBD} (Table~\ref{tab:galmodgrid}), we do find a higher scale height than that of M dwarfs, as \citet{rya05} found, of the order of $h_{z,thin} \sim 450 pc$. Nonetheless, there is a degeneracy between the thin disk scale height and $n_{sun,thick}$ and therefore we cannot yet rule out a models with $h_{z,thin}=300 \ pc$: considering the error in the number counts $\sigma_{LT} \sim 900$, there are models of $h_{z,thin}=300 \ pc$ which predicts a number of LT types within $1\sigma$ of our measurement. In Fig.~\ref{fig:scaleheight} we compare the number counts to various models taken from \textit{GalmodBD}. In Grey we show two models for $h_{z,thin}=300 \ pc$, one with $n_{sun,thick} = 5\%$ (lower limit) and another with $n_{sun,thick} = 20\%$ (upper limit). In dashed blue we show models for $h_{z,thin}=400 \ pc$, for same values of $n_{sun,thick}$ and in red points we show models for $h_{z,thin}=500 \ pc$.

In order to constraint the thin disk scale height, we will perform a Markov chain Monte Carlo analysis which will marginalize over $n_{sun,thick}$. This will be the scope of a future analysis. We will wait until the full coverage of DES and VHS have been achieved, covering $\sim 5000 deg^{2}$. Here we will apply the same methodology presented here, for magnitudes $z<22.7$, increasing both the area and the distance of the surveyed sample. 

\section{Conclusion}
\label{sec:conclusions}

In this paper, we apply a photometric brown dwarf classification scheme based on \citet{skr14} to $DES \cap VHS \cap AllWISE$ data using 8 bands covering a wavelength range between 0.7 $\mu m$ and 4.6 $\mu m$. Since several degeneracies are found in colour space between spectral types M,L and T, the use of multiple bands are required for a good spectral type calibration.

In comparison with \citet{skr14}, we can go to greater distances in the L regime due to the deeper DES and VHS samples, in contrast with the $SDSS \cap UKIDSS$ design. This way, we classify 11,745 BDs in the spectral regime from L0 to T9 in $\approx 2,400 \ deg^{2}$, with a similar spectral resolution of 1 spectral type. We make this catalogue public in electronic format. It can be downloaded from \url{https://des.ncsa.illinois.edu/releases/other/y3-mlt}. This is the largest LT sample ever published. We further estimate the purity and completeness of the sample. We estimate the sample to be $\sim 77\%$ complete and $\sim 74\%$ pure at $z_{AB}<22$. Finally, we can calculate the total number of LTs in the $DES \cap VHS$ footprint to be $\sim 12,300 \pm 900$. 

During the classification, we identify 2,818 possible extragalactic sources that we removed from the catalogue, containing 57 quasars at high redshift. The removal of extragalactic sources increases the uncertainty in the determination of the number of LT types.

In parallel, we have presented the \textit{GalmodBD} simulation, a simulation that computes LT number counts as a function of SFH parameters. In our analysis, we found that the thin disk scale height and the thin-to-thick disk normalisation were the parameters that most affect the number counts. Nonetheless, more free parameters are available in the simulation. When comparing the simulation output with the number of LT expected in the footprint, we put constraints on the thin disk scale height for early L-types, finding a value that is in agreement with recent measurements, like the one found in \citet{Sor18} with $h_{z} \sim 400 \ pc$.  

Having these two ingredients, a robust simulation of number counts, and a methodology to select BDs in $DES \cap VHS \cap AllWISE$ will allow us to do a precise measurement of the thin disk scale height of the L population, putting brown dwarf science in its Galactic context. This will be the scope of future analyses.

\section*{Acknowledgments}

Funding for the DES Projects has been provided by the U.S. Department of Energy, the U.S. National Science Foundation, the Ministry of Science and Education of Spain, 
the Science and Technology Facilities Council of the United Kingdom, the Higher Education Funding Council for England, the National Center for Supercomputing 
Applications at the University of Illinois at Urbana-Champaign, the Kavli Institute of Cosmological Physics at the University of Chicago, 
the Center for Cosmology and Astro-Particle Physics at the Ohio State University,
the Mitchell Institute for Fundamental Physics and Astronomy at Texas A\&M University, Financiadora de Estudos e Projetos, 
Funda{\c c}{\~a}o Carlos Chagas Filho de Amparo {\`a} Pesquisa do Estado do Rio de Janeiro, Conselho Nacional de Desenvolvimento Cient{\'i}fico e Tecnol{\'o}gico and 
the Minist{\'e}rio da Ci{\^e}ncia, Tecnologia e Inova{\c c}{\~a}o, the Deutsche Forschungsgemeinschaft and the Collaborating Institutions in the Dark Energy Survey. 

The Collaborating Institutions are Argonne National Laboratory, the University of California at Santa Cruz, the University of Cambridge, Centro de Investigaciones Energ{\'e}ticas, 
Medioambientales y Tecnol{\'o}gicas-Madrid, the University of Chicago, University College London, the DES-Brazil Consortium, the University of Edinburgh, 
the Eidgen{\"o}ssische Technische Hochschule (ETH) Z{\"u}rich, 
Fermi National Accelerator Laboratory, the University of Illinois at Urbana-Champaign, the Institut de Ci{\`e}ncies de l'Espai (IEEC/CSIC), 
the Institut de F{\'i}sica d'Altes Energies, Lawrence Berkeley National Laboratory, the Ludwig-Maximilians Universit{\"a}t M{\"u}nchen and the associated Excellence Cluster Universe, 
the University of Michigan, the National Optical Astronomy Observatory, the University of Nottingham, The Ohio State University, the University of Pennsylvania, the University of Portsmouth, 
SLAC National Accelerator Laboratory, Stanford University, the University of Sussex, Texas A\&M University, and the OzDES Membership Consortium.

Based in part on observations at Cerro Tololo Inter-American Observatory, National Optical Astronomy Observatory, which is operated by the Association of 
Universities for Research in Astronomy (AURA) under a cooperative agreement with the National Science Foundation.

The DES data management system is supported by the National Science Foundation under Grant Numbers AST-1138766 and AST-1536171.
The DES participants from Spanish institutions are partially supported by MINECO under grants AYA2015-71825, ESP2015-66861, FPA2015-68048, SEV-2016-0588, SEV-2016-0597, and MDM-2015-0509, 
some of which include ERDF funds from the European Union. IFAE is partially funded by the CERCA program of the Generalitat de Catalunya.
Research leading to these results has received funding from the European Research
Council under the European Union's Seventh Framework Program (FP7/2007-2013) including ERC grant agreements 240672, 291329, and 306478.
We  acknowledge support from the Australian Research Council Centre of Excellence for All-sky Astrophysics (CAASTRO), through project number CE110001020, and the Brazilian Instituto Nacional de Ci\^encia
e Tecnologia (INCT) e-Universe (CNPq grant 465376/2014-2).

This manuscript has been authored by Fermi Research Alliance, LLC under Contract No. DE-AC02-07CH11359 with the U.S. Department of Energy, Office of Science, Office of High Energy Physics. The United States Government retains and the publisher, by accepting the article for publication, acknowledges that the United States Government retains a non-exclusive, paid-up, irrevocable, world-wide license to publish or reproduce the published form of this manuscript, or allow others to do so, for United States Government purposes.

This publication makes use of data products from the Wide-field Infrared Survey Explorer, which is a joint project of the University of California, Los Angeles, and the Jet Propulsion Laboratory/California Institute of Technology, and NEOWISE, which is a project of the Jet Propulsion Laboratory/California Institute of Technology. WISE and NEOWISE are funded by the National Aeronautics and Space Administration. 

The analysis presented here is based on observations obtained
as part of the VISTA Hemisphere Survey, ESO Programme, 179.A-2010 (PI: McMahon). 

This paper has gone through internal review by the DES collaboration.

ACR acknowledges financial support provided by the PAPDRJ CAPES/FAPERJ Fellowship and by ``Unidad de Excelencia Mar\'ia de Maeztu de CIEMAT - F\'isica de Part\'iculas (Proyecto MDM)".

The authors would like to thank Jacqueline Faherty and Joe Filippazzo for their help on the early stages of this project. 

\bibliography{references} 

\begin{thebibliography}{78}
\expandafter\ifx\csname natexlab\endcsname\relax\def\natexlab#1{#1}\fi

\bibitem[Abell et~al.(2009)Abell, Allison, Anderson et~al.]{abell2009lsst}
Abell P.~A., Allison J., Anderson S.~F., et~al., 2009

\bibitem[Aihara et~al.(2018)Aihara, Armstrong, Bickerton et~al.]{Aih18}
Aihara H., Armstrong R., Bickerton S., et~al., 2018, Publication of the
  Astronomical Society of Japan, 70, Special Issue 1

\bibitem[Albert et~al.(2011)Albert, Artigau, Delorme et~al.]{albert2011}
Albert L., Artigau {\'E}., Delorme P., et~al., 2011, Aj, 141, 203

\bibitem[An et~al.(2008)An, Johnson, Clem et~al.]{Ann08}
An D., Johnson J.~A., Clem J.~L., et~al., 2008, The Astrophysical Journal
  Supplement Series, 179, 2, 326

\bibitem[{Arnouts} et~al.(1999){Arnouts}, {Cristiani}, {Moscardini}
  et~al.]{Arn99}
{Arnouts} S., {Cristiani} S., {Moscardini} L., et~al., 1999, \mnras, 310, 540

\bibitem[{Arnouts} et~al.(2007){Arnouts}, {Walcher}, {Le F{\`e}vre}
  et~al.]{arn07}
{Arnouts} S., {Walcher} C.~J., {Le F{\`e}vre} O., et~al., 2007, \aap, 476, 137

\bibitem[{Banerji} et~al.(2015){Banerji}, {Jouvel}, {Lin} et~al.]{bane15}
{Banerji} M., {Jouvel} S., {Lin} H., et~al., 2015, \mnras, 446, 2523

\bibitem[{Basri} \& {Reiners}(2006)]{Basri2006}
{Basri} G., {Reiners} A., 2006, \aj, 132, 663

\bibitem[{Bertin} \& {Arnouts}(1996)]{Ber99}
{Bertin} E., {Arnouts} S., 1996, \aaps, 117, 393

\bibitem[{Bochanski} et~al.(2007){Bochanski}, {Munn}, {Hawley}, {West}, {Covey}
  \& {Schneider}]{boch07}
{Bochanski} J.~J., {Munn} J.~A., {Hawley} S.~L., {West} A.~A., {Covey} K.~R.,
  {Schneider} D.~P., 2007, \aj, 134, 2418

\bibitem[{Bruzual} \& {Charlot}(2003)]{bru03}
{Bruzual} G., {Charlot} S., 2003, \mnras, 344, 1000

\bibitem[Burgasser(2007)]{bur07}
Burgasser A.~J., 2007, ApJ, 659, 655

\bibitem[Burgasser et~al.(2006{\natexlab{a}})Burgasser, Geballe, Leggett,
  Kirkpatrick \& Golimowski]{bur06b}
Burgasser A.~J., Geballe T.~R., Leggett S.~K., Kirkpatrick J.~D., Golimowski
  D.~A., 2006{\natexlab{a}}, ApJ, 637, 1067

\bibitem[Burgasser et~al.(2003)Burgasser, Kirkpatrick, Burrows et~al.]{bur03}
Burgasser A.~J., Kirkpatrick J.~D., Burrows A., et~al., 2003, ApJ, 592, 1186

\bibitem[Burgasser et~al.(2006{\natexlab{b}})Burgasser, Kirkpatrick, Cruz
  et~al.]{bur06c}
Burgasser A.~J., Kirkpatrick J.~D., Cruz K.~L., et~al., 2006{\natexlab{b}},
  ApJSupplement, 166, 585

\bibitem[{Burke} et~al.(2018){Burke}, {Rykoff}, {Allam} et~al.]{burk18}
{Burke} D.~L., {Rykoff} E.~S., {Allam} S., et~al., 2018, \aj, 155, 41

\bibitem[{Burningham}(2018)]{bur18}
{Burningham} B., 2018, ArXiv e-prints

\bibitem[Burningham et~al.(2013)Burningham, Cardoso, Smith et~al.]{Bur13}
Burningham B., Cardoso C.~V., Smith L., et~al., 2013, Monthly Notices of the
  Royal Astronomical Society, 433, 1, 457

\bibitem[{Burningham} et~al.(2010){Burningham}, {Pinfield}, {Lucas}
  et~al.]{bur10}
{Burningham} B., {Pinfield} D.~J., {Lucas} P.~W., et~al., 2010, \mnras, 406,
  1885

\bibitem[{Burstein} \& {Heiles}(1982)]{Bur82}
{Burstein} D., {Heiles} C., 1982, Aj, 87, 1165

\bibitem[Chiu et~al.(2006)Chiu, Fan, Leggett et~al.]{chiu2006}
Chiu K., Fan X., Leggett S.~K., et~al., 2006, Aj, 131, 2722

\bibitem[{Coleman} et~al.(1980){Coleman}, {Wu} \& {Weedman}]{col80}
{Coleman} G.~D., {Wu} C.-C., {Weedman} D.~W., 1980, \apjs, 43, 393

\bibitem[{Cross} et~al.(2012){Cross}, {Collins}, {Mann} et~al.]{cross12}
{Cross} N.~J.~G., {Collins} R.~S., {Mann} R.~G., et~al., 2012, \aap, 548, A119

\bibitem[{Cruz} et~al.(2007){Cruz}, {Reid}, {Kirkpatrick} et~al.]{Cru07}
{Cruz} K.~L., {Reid} I.~N., {Kirkpatrick} J.~D., et~al., 2007, Aj, 133, 439

\bibitem[{Dalton} et~al.(2006){Dalton}, {Caldwell}, {Ward} et~al.]{dalton06}
{Dalton} G.~B., {Caldwell} M., {Ward} A.~K., et~al., 2006, in { Society of
  Photo-Optical Instrumentation Engineers (SPIE) Conference Series\/}, vol.
  6269 of { \procspie\/},  62690X

\bibitem[{Dark Energy Survey Collaboration} et~al.(2016){Dark Energy Survey
  Collaboration}, {Abbott}, {Abdalla} et~al.]{des16}
{Dark Energy Survey Collaboration}, {Abbott} T., {Abdalla} F.~B., et~al., 2016,
  \mnras, 460, 1270

\bibitem[{Dark Energy Survey Collaboration} et~al.(2018){Dark Energy Survey
  Collaboration}, {Abbott}, {Abdalla} et~al.]{abb18}
{Dark Energy Survey Collaboration}, {Abbott} T.~M.~C., {Abdalla} F.~B., et~al.,
  2018, \apjs, 239, 18

\bibitem[Day-Jones et~al.(2013)Day-Jones, Marocco, Pinfield et~al.]{DJ13}
Day-Jones A.~C., Marocco F., Pinfield D.~J., et~al., 2013, MNRAS, 430, 1171

\bibitem[{Drlica-Wagner} et~al.(2018){Drlica-Wagner}, {Sevilla-Noarbe},
  {Rykoff} et~al.]{drl18}
{Drlica-Wagner} A., {Sevilla-Noarbe} I., {Rykoff} E.~S., et~al., 2018, The
  Astrophysical Journal Supplement Series, 235, 33

\bibitem[{Dupuy} \& {Liu}(2012)]{Dup12}
{Dupuy} T.~J., {Liu} M.~C., 2012, Apjs, 201, 19

\bibitem[Faherty et~al.(2009)Faherty, Burgasser, Cruz, Shara, Walter \&
  Gelino]{fah09}
Faherty J.~K., Burgasser A.~J., Cruz K.~L., Shara M.~M., Walter F.~M., Gelino
  C.~R., 2009, Aj, 137, 1

\bibitem[Faherty et~al.(2012)Faherty, Burgasser, Walter et~al.]{fah12}
Faherty J.~K., Burgasser A.~J., Walter F.~M., et~al., 2012, ApJ, 752, 56

\bibitem[{Faherty} et~al.(2016){Faherty}, {Riedel}, {Cruz} et~al.]{faherty2016}
{Faherty} J.~K., {Riedel} A.~R., {Cruz} K.~L., et~al., 2016, \apjs, 225, 10

\bibitem[Folkes et~al.(2007)Folkes, Pinfield, Kendall \& Jones]{fol07}
Folkes S.~L., Pinfield D.~J., Kendall T.~R., Jones H.~R.~A., 2007, MNRAS, 378,
  901

\bibitem[{Fontanive} et~al.(2018){Fontanive}, {Biller}, {Bonavita} \&
  {Allers}]{font18}
{Fontanive} C., {Biller} B., {Bonavita} M., {Allers} K., 2018, \mnras, 479,
  2702

\bibitem[{Gagn{\'e}} et~al.(2017){Gagn{\'e}}, {Faherty}, {Mamajek}
  et~al.]{gagne2017}
{Gagn{\'e}} J., {Faherty} J.~K., {Mamajek} E.~E., et~al., 2017, \apjs, 228, 18

\bibitem[{Gaia Collaboration} et~al.(2018){Gaia Collaboration}, {Brown},
  {Vallenari} et~al.]{gaia}
{Gaia Collaboration}, {Brown} A.~G.~A., {Vallenari} A., et~al., 2018, \aap,
  616, A1

\bibitem[{Ilbert} et~al.(2006){Ilbert}, {Arnouts}, {McCracken} et~al.]{Ilb06}
{Ilbert} O., {Arnouts} S., {McCracken} H.~J., et~al., 2006, \aap, 457, 841

\bibitem[Juri{\'c} et~al.(2008)Juri{\'c}, Ivezi{\'c}, Brooks et~al.]{jur08}
Juri{\'c} M., Ivezi{\'c} {\v Z}., Brooks A., et~al., 2008, \apj, 673, 864

\bibitem[{Kerber} et~al.(2001){Kerber}, {Javiel} \& {Santiago}]{Ker01}
{Kerber} L.~O., {Javiel} S.~C., {Santiago} B.~X., 2001, \aap, 365, 424

\bibitem[{Kinney} et~al.(1996){Kinney}, {Calzetti}, {Bohlin}, {McQuade},
  {Storchi-Bergmann} \& {Schmitt}]{kin96}
{Kinney} A.~L., {Calzetti} D., {Bohlin} R.~C., {McQuade} K., {Storchi-Bergmann}
  T., {Schmitt} H.~R., 1996, \apj, 467, 38

\bibitem[Kirkpatrick et~al.(2011)Kirkpatrick, Cushing, Gelino et~al.]{kirk2011}
Kirkpatrick J.~D., Cushing M.~C., Gelino C.~R., et~al., 2011, ApJs, 197, 19

\bibitem[Kirkpatrick et~al.(1999)Kirkpatrick, Reid, Liebert et~al.]{kir99}
Kirkpatrick J.~D., Reid I.~N., Liebert J., et~al., 1999, ApJ, 519, 802

\bibitem[Lawrence et~al.(2007)Lawrence, Warren, Almaini et~al.]{lawrence07}
Lawrence A., Warren S., Almaini O., et~al., 2007, Monthly Notices of the Royal
  Astronomical Society, 379, 4, 1599

\bibitem[{Leggett} et~al.(2015){Leggett}, {Morley}, {Marley} \&
  {Saumon}]{leggett2015}
{Leggett} S.~K., {Morley} C.~V., {Marley} M.~S., {Saumon} D., 2015, \apj, 799,
  37

\bibitem[{Leggett} et~al.(2013){Leggett}, {Morley}, {Marley}, {Saumon},
  {Fortney} \& {Visscher}]{leggett2013}
{Leggett} S.~K., {Morley} C.~V., {Marley} M.~S., {Saumon} D., {Fortney} J.~J.,
  {Visscher} C., 2013, \apj, 763, 130

\bibitem[{Liu} et~al.(2013){Liu}, {Magnier}, {Deacon} et~al.]{liu2013}
{Liu} M.~C., {Magnier} E.~A., {Deacon} N.~R., et~al., 2013, \apjl, 777, L20

\bibitem[Lodieu et~al.(2012)Lodieu, Burningham, Day-Jones et~al.]{lodieu2012}
Lodieu N., Burningham B., Day-Jones A., et~al., 2012, AaP, 548, A53

\bibitem[Looper et~al.(2008)Looper, Kirkpatrick, Cutri et~al.]{loo08}
Looper D.~L., Kirkpatrick J.~D., Cutri R.~M., et~al., 2008, ApJ, 686, 528

\bibitem[Luhman(2012)]{luh12}
Luhman K.~L., 2012, Araa, 50, 65

\bibitem[{Mainzer} et~al.(2011){Mainzer}, {Bauer}, {Grav} et~al.]{mai11}
{Mainzer} A., {Bauer} J., {Grav} T., et~al., 2011, \apj, 731, 53

\bibitem[{Marocco} et~al.(2014){Marocco}, {Day-Jones}, {Lucas} et~al.]{pec14}
{Marocco} F., {Day-Jones} A.~C., {Lucas} P.~W., et~al., 2014, \mnras, 439, 372

\bibitem[Marocco et~al.(2015)Marocco, Jones, Day-Jones et~al.]{Mar15}
Marocco F., Jones H. R.~A., Day-Jones A.~C., et~al., 2015, Monthly Notices of
  the Royal Astronomical Society, 449, 4, 3651

\bibitem[Marocco et~al.(2017)Marocco, Pinfield, Cook et~al.]{mar17}
Marocco F., Pinfield D.~J., Cook N.~J., et~al., 2017, MNRAS, 470, 4885

\bibitem[{Maxted} \& {Jeffries}(2005)]{Maxted2005}
{Maxted} P.~F.~L., {Jeffries} R.~D., 2005, \mnras, 362, L45

\bibitem[{McMahon} et~al.(2013){McMahon}, {Banerji}, {Gonzalez}
  et~al.]{mcmahon13}
{McMahon} R.~G., {Banerji} M., {Gonzalez} E., et~al., 2013, The Messenger, 154,
  35

\bibitem[{Miyazaki} et~al.(2018){Miyazaki}, {Komiyama}, {Kawanomoto}
  et~al.]{Miy18}
{Miyazaki} S., {Komiyama} Y., {Kawanomoto} S., et~al., 2018, \pasj, 70, S1

\bibitem[{Morganson} et~al.(2018){Morganson}, {Gruendl}, {Menanteau}
  et~al.]{mor18}
{Morganson} E., {Gruendl} R.~A., {Menanteau} F., et~al., 2018, \pasp, 130, 7,
  074501

\bibitem[Pinfield et~al.(2006)Pinfield, Jones, Lucas et~al.]{pin06}
Pinfield D.~J., Jones H.~R.~A., Lucas P.~W., et~al., 2006, MNRAS, 368, 1281

\bibitem[{Prevot} et~al.(1984){Prevot}, {Lequeux}, {Maurice}, {Prevot} \&
  {Rocca-Volmerange}]{pre84}
{Prevot} M.~L., {Lequeux} J., {Maurice} E., {Prevot} L., {Rocca-Volmerange} B.,
  1984, \aap, 132, 389

\bibitem[Reed et~al.(2015)Reed, McMahon, Banerji et~al.]{ree15}
Reed S.~L., McMahon R.~G., Banerji M., et~al., 2015, MNRAS, 454, 3952

\bibitem[Reed et~al.(2017)Reed, McMahon, Martini et~al.]{ree17}
Reed S.~L., McMahon R.~G., Martini P., et~al., 2017, MNRAS, 468, 4702

\bibitem[Ryan et~al.(2005)Ryan, Hathi, Cohen \& Windhorst]{rya05}
Ryan Jr. R.~E., Hathi N.~P., Cohen S.~H., Windhorst R.~A., 2005, ApJLetters,
  631, L159

\bibitem[{Santiago} et~al.(2010){Santiago}, {Yanny} \& {Yanny}]{San10}
{Santiago} B., {Yanny} B., {Yanny}, 2010, in { Stellar Populations - Planning
  for the Next Decade\/}, edited by G.~R. {Bruzual}, S.~{Charlot}, vol. 262 of
  { IAU Symposium\/},  265--269

\bibitem[{Santiago} et~al.(1996){Santiago}, {Gilmore} \& {Elson}]{San96}
{Santiago} B.~X., {Gilmore} G., {Elson} R.~A.~W., 1996, \mnras, 281, 871

\bibitem[{Schlafly} \& {Finkbeiner}(2011)]{sch11}
{Schlafly} E.~F., {Finkbeiner} D.~P., 2011, \apj, 737, 103

\bibitem[{Schlegel} et~al.(1998){Schlegel}, {Finkbeiner} \& {Davis}]{Sch98}
{Schlegel} D.~J., {Finkbeiner} D.~P., {Davis} M., 1998, Apj, 500, 525

\bibitem[Schmidt et~al.(2010)Schmidt, West, Hawley \& Pineda]{sch10}
Schmidt S.~J., West A.~A., Hawley S.~L., Pineda J.~S., 2010, Aj, 139, 1808

\bibitem[{Skemer} et~al.(2016){Skemer}, {Morley}, {Allers} et~al.]{skemer2016}
{Skemer} A.~J., {Morley} C.~V., {Allers} K.~N., et~al., 2016, \apjl, 826, L17

\bibitem[{Skrzypek} et~al.(2016){Skrzypek}, {Warren} \& {Faherty}]{skr2016}
{Skrzypek} N., {Warren} S.~J., {Faherty} J.~K., 2016, \aap, 589, A49

\bibitem[Skrzypek et~al.(2015)Skrzypek, Warren, Faherty, Mortlock, Burgasser \&
  Hewett]{skr14}
Skrzypek N., Warren S.~J., Faherty J.~K., Mortlock D.~J., Burgasser A.~J.,
  Hewett P.~C., 2015, \aap, 574, A78

\bibitem[Smith et~al.(2014)Smith, Lucas, Burningham et~al.]{smi14}
Smith L., Lucas P.~W., Burningham B., et~al., 2014, MNRAS, 437, 3603

\bibitem[{Sorahana} et~al.(2018){Sorahana}, {Nakajima} \& {Matsuoka}]{Sor18}
{Sorahana} S., {Nakajima} T., {Matsuoka} Y., 2018, ArXiv e-prints

\bibitem[Tie et~al.(2017)Tie, Martini, Mudd et~al.]{tie17}
Tie S.~S., Martini P., Mudd D., et~al., 2017, Aj, 153, 107

\bibitem[{Warren} et~al.(2007){Warren}, {Cross}, {Dye} et~al.]{war07}
{Warren} S.~J., {Cross} N.~J.~G., {Dye} S., et~al., 2007, ArXiv Astrophysics
  e-prints

\bibitem[{West} et~al.(2011){West}, {Morgan}, {Bochanski}, {Andersen} \& {et
  al.}]{west11}
{West} A.~A., {Morgan} D.~P., {Bochanski} J.~J., {Andersen} J.~M., {et al.},
  2011, aj, 141, 97

\bibitem[{Wright} et~al.(2010){Wright}, {Eisenhardt}, {Mainzer} et~al.]{wri10}
{Wright} E.~L., {Eisenhardt} P.~R.~M., {Mainzer} A.~K., et~al., 2010, \aj, 140,
  1868

\bibitem[{Zhang} et~al.(2017){Zhang}, {Homeier}, {Pinfield} et~al.]{zhang2017}
{Zhang} Z.~H., {Homeier} D., {Pinfield} D.~J., et~al., 2017, \mnras, 468, 261

\end{thebibliography}

\section{Affiliations}

$^{1}$ Centro de Investigaciones Energ\'eticas, Medioambientales y Tecnol\'ogicas (CIEMAT), Madrid, Spain\\
$^{2}$ Laborat\'orio Interinstitucional de e-Astronomia - LIneA, Rua Gal. Jos\'e Cristino 77, Rio de Janeiro, RJ - 20921-400, Brazil\\
$^{3}$ Observat\'orio Nacional, Rua Gal. Jos\'e Cristino 77, Rio de Janeiro, RJ - 20921-400, Brazil\\
$^{4}$ Instituto de F\'\i sica, UFRGS, Caixa Postal 15051, Porto Alegre, RS - 91501-970, Brazil\\
$^{5}$ Center for Astrophysics Research, University of Hertfordshire, Hatfield AL10 9AB, UK\\
$^{6}$ Harvard-Smithsonian Center for Astrophysics, Cambridge, MA 02138, USA\\
$^{7}$ Black Hole Initiative at Harvard University, 20 Garden Street, Cambridge, MA 02138, USA\\
$^{8}$ George P. and Cynthia Woods Mitchell Institute for Fundamental Physics and Astronomy, and Department of Physics and Astronomy, Texas A\&M University, College Station, TX 77843,  USA\\
$^{9}$ Institute of Astronomy, University of Cambridge, Madingley Road, Cambridge CB3 0HA, UK\\
$^{10}$ Kavli Institute for Cosmology, University of Cambridge, Madingley Road, Cambridge CB3 0HA, UK\\
$^{11}$ Physics Department, 2320 Chamberlin Hall, University of Wisconsin-Madison, 1150 University Avenue Madison, WI  53706-1390\\
$^{12}$ LSST, 933 North Cherry Avenue, Tucson, AZ 85721, USA\\
$^{13}$ Fermi National Accelerator Laboratory, P. O. Box 500, Batavia, IL 60510, USA\\
$^{14}$ Cerro Tololo Inter-American Observatory, National Optical Astronomy Observatory, Casilla 603, La Serena, Chile\\
$^{15}$ Instituto de Fisica Teorica UAM/CSIC, Universidad Autonoma de Madrid, 28049 Madrid, Spain\\
$^{16}$ Department of Physics and Astronomy, University of Pennsylvania, Philadelphia, PA 19104, USA\\
$^{17}$ Department of Physics \& Astronomy, University College London, Gower Street, London, WC1E 6BT, UK\\
$^{18}$ Kavli Institute for Particle Astrophysics \& Cosmology, P. O. Box 2450, Stanford University, Stanford, CA 94305, USA\\
$^{19}$ SLAC National Accelerator Laboratory, Menlo Park, CA 94025, USA\\
$^{20}$ Department of Astronomy, University of Illinois at Urbana-Champaign, 1002 W. Green Street, Urbana, IL 61801, USA\\
$^{21}$ National Center for Supercomputing Applications, 1205 West Clark St., Urbana, IL 61801, USA\\
$^{22}$ Institut de F\'{\i}sica d'Altes Energies (IFAE), The Barcelona Institute of Science and Technology, Campus UAB, 08193 Bellaterra (Barcelona) Spain\\
$^{23}$ Kavli Institute for Cosmological Physics, University of Chicago, Chicago, IL 60637, USA\\
$^{24}$ Institut d'Estudis Espacials de Catalunya (IEEC), 08034 Barcelona, Spain\\
$^{25}$ Institute of Space Sciences (ICE, CSIC),  Campus UAB, Carrer de Can Magrans, s/n,  08193 Barcelona, Spain\\
$^{26}$ Santa Cruz Institute for Particle Physics, Santa Cruz, CA 95064, USA\\
$^{27}$ Instituci\'o Catalana de Recerca i Estudis Avan\c{c}ats, E-08010 Barcelona, Spain\\
$^{28}$ Department of Astrophysical Sciences, Princeton University, Peyton Hall, Princeton, NJ 08544, USA\\
$^{29}$ School of Physics and Astronomy, University of Southampton,  Southampton, SO17 1BJ, UK\\
$^{30}$ Instituto de F\'isica Gleb Wataghin, Universidade Estadual de Campinas, 13083-859, Campinas, SP, Brazil\\
$^{31}$ Computer Science and Mathematics Division, Oak Ridge National Laboratory, Oak Ridge, TN 37831\\
$^{32}$ Department of Physics, University of Michigan, Ann Arbor, MI 48109, USA\\
$^{33}$ Argonne National Laboratory, 9700 South Cass Avenue, Lemont, IL 60439, USA

\appendix

\section{Lephare configuration}
\label{app:lephare}

In this section we detail the \verb+Lephare+ configuration we used to separate extragalactic sources from the main MLT catalogue. We run separately \verb+Lephare+ for galaxy and quasar templates.

We used \verb+AVEROIN+ galaxy template library, containing 62 different templates including starburst, spiral, elliptical and irregular galaxies from \citet{col80}, \citet{kin96}, and \citet{bru03} and tuned in the mid-infrared (from 3.6 to 4.5 microns) based on VVDS-CFHTLS-SWIRE photometry and VVDS spectra. They were used for the first in ~\citet{arn07}. We allow for internal extinction following the extinction law by \citet{pre84} up to a E(B-V) = 0.3. In Table~\ref{tab:lephareconf} we summarize the configuration parameters we used in \verb+Lephare+.

For quasars, we use the default quasar template list from \verb+Lephare+ and we refer to their documentation for details about the templates used. In Table~\ref{tab:lephareconfqso} we summarize the configuration used for the quasar run. 

We further need the filter passbands for $i,z,Y$ (DES), $J,H,K_{s}$ (VHS) and $W1,W2$ (WISE). The DES passbands are the most updated versions of the calibrated transmission curves as shown in \citet{burk18}, for the VHS passbands, we use the curves given at \verb+ESO+ instrument description page\footnote{\url{http://www.eso.org/sci/facilities/paranal/instruments/vircam/inst.html}} while for the WISE filters we used the curves given at the WISE documentation page\footnote{\url{http://www.astro.ucla.edu/~wright/WISE/passbands.html}}.

Finally, we add an error of 0.07 in quadrature to each magnitude error as we did for the main \textit{classif} run and transform magnitudes to AB previous to run \verb+Lephare+.

\begin{table*}
\centering
\caption{Lephare configuration used when fitting the data to galaxy templates.}
\label{tab:lephareconf}
\begin{tabular}{c c c}
NAME & VALUE & EXPLANATION \\
\hline
\verb+GAL_SED+ & AVEROIN\_MOD.list & List of 62 galaxy templates \\
\verb+GAL_FSCALE+ & 1. & Arbitrary Flux Scale \\
\verb+FILTER_LIST+ & $i,z,Y,J,H,K_{s},W1,W2$ & List of paths to ascii files with passbands \\
\verb+TRANS_TYPE+  &     0 & Transmission type (0 = Energy, 1 = Nb of photons) \\
\verb+FILTER_CALIB+ & 0 & Filter calibration (0 = \verb+fnu=ctt+) \\
\verb+MAGTYPE+ & AB & Magnitude type (AB or VEGA) \\
\verb+Z_STEP+  & 0.04, 6., 0.1 & dz, zmax, dzsup \\
\verb+COSMOLOGY+ & 70, 0.3, 0.7 & H0, $\Omega_{M}$, $\Omega_{\Lambda}$ \\
\verb+EXTINC_LAW+ & SMC\_prevot.dat  & Extinction law \\
\verb+EB_V+ & 0., 0.05, 0.1, 0.15, 0.2, 0.3 & Allowed E(B-V) values \\
\verb+MOD_EXTINC+ & 38, 62 & Templates for which to apply extinction law \\
\verb+EM_LINES+  & NO & Allow emission lines \\
\verb+CAT_MAG+ & AB & Input Magnitude (AB or VEGA) \\
\verb+ERR_SCALE+ & 0.07 & Errors per band added in quadrature \\
\verb+Z_INTERP+ & YES & Redshift interpolation \\
\end{tabular}
\end{table*}

\begin{table*}
\centering
\caption{Lephare configuration used to fit the data to quasar templates.}
\label{tab:lephareconfqso}
\begin{tabular}{c c c}
NAME & VALUE & EXPLANATION \\
\hline
\verb+GAL_SED+ & QSO\_MOD.list & List of 28 quasar templates \\
\verb+GAL_FSCALE+ & 1. & Arbitrary Flux Scale \\
\verb+FILTER_LIST+ & $i,z,Y,J,H,K_{s},W1,W2$ & List of paths to ascii files with passbands \\
\verb+TRANS_TYPE+  &     0 & Transmission type (0 = Energy, 1 = Nb of photons) \\
\verb+FILTER_CALIB+ & 0 & Filter calibration (0 = \verb+fnu=ctt+) \\
\verb+MAGTYPE+ & AB & Magnitude type (AB or VEGA) \\
\verb+Z_STEP+  & 0.04, 9., 0.1 & dz, zmax, dzsup \\
\verb+COSMOLOGY+ & 70, 0.3, 0.7 & H0, $\Omega_{M}$, $\Omega_{\Lambda}$ \\
\verb+EB_V+ & None & Allowed E(B-V) values \\
\verb+CAT_MAG+ & AB & Input Magnitude (AB or VEGA) \\
\verb+ERR_SCALE+ & 0.07 & Errors per band added in quadrature \\
\verb+Z_INTERP+ & YES & Redshift interpolation \\
\end{tabular}
\end{table*}

\section{Galaxy contamination in the MLT catalogue}
\label{app:lepharesg}

In this section we describe the extragalactic population found in the colour space defined in Section~\ref{sec:targetselection} after running \verb+Lephare+. From the list of 2818 targets, 2,761 are galaxies, and 57 are quasars. Next, we show the properties of the galaxy population.

\begin{figure}
\includegraphics[width=\columnwidth]{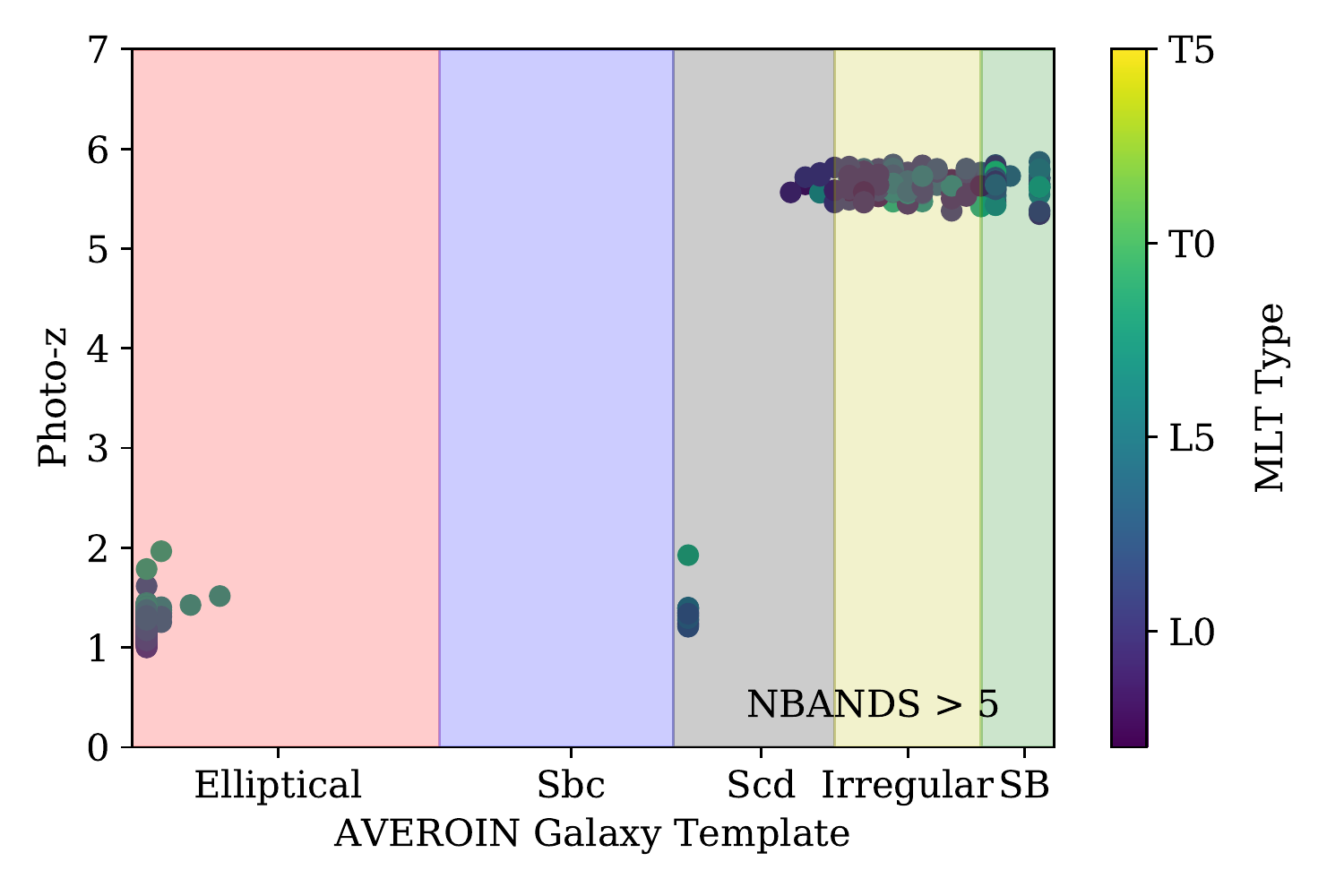}
\caption{Best galaxy templates versus redshift for extragalactic targets with $NBANDS>5$. The colour scale is given by the colour bar and represent the best-fit MLT type.}
\label{fig:modvsphotoz}
\end{figure}

\begin{figure}
 \includegraphics[width=\columnwidth]{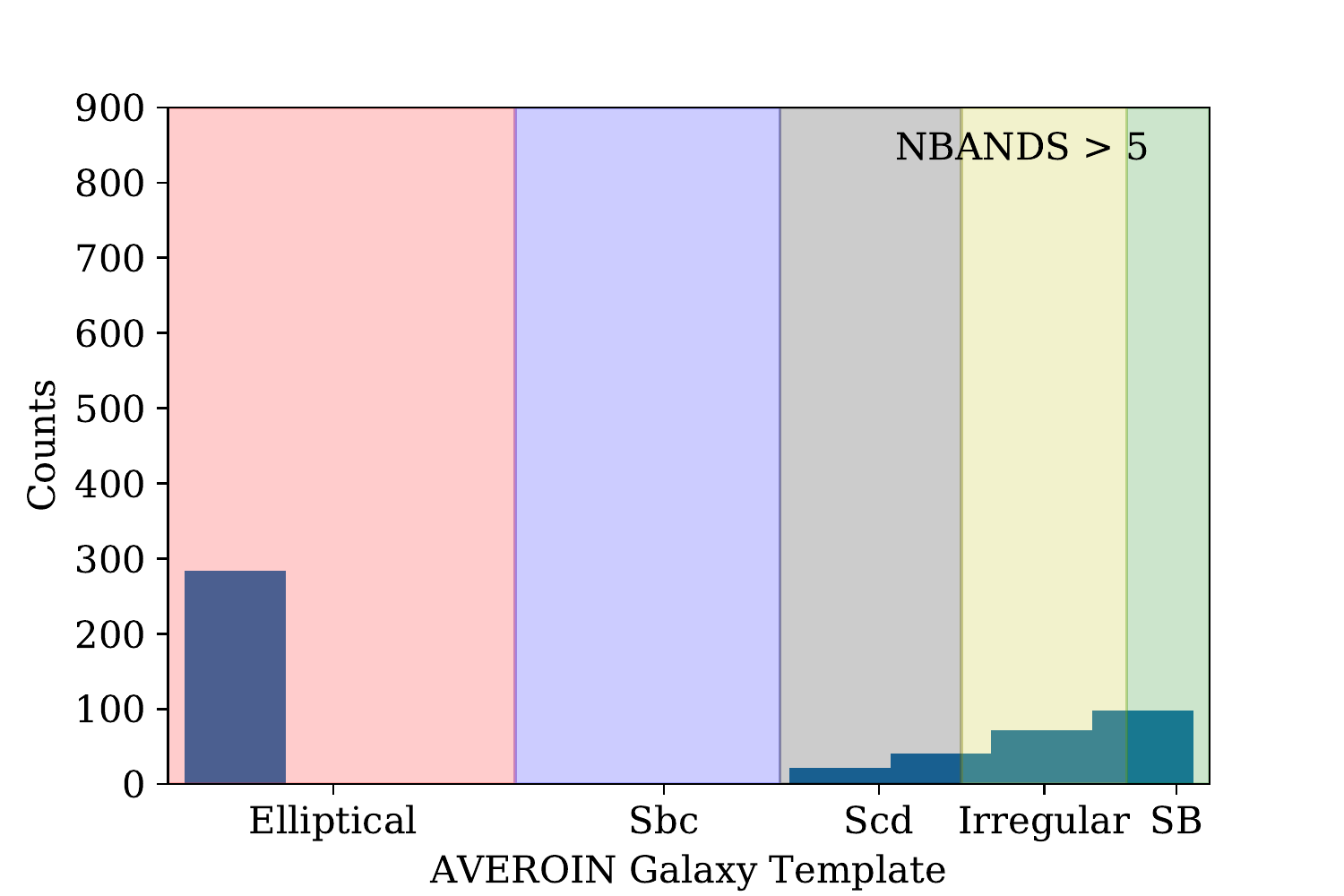}
 \caption{Best galaxy template distribution for extragalactic targets with $NBANDS>5$.}
  \label{fig:nmod}
\end{figure}

To avoid biases in our conclusions due to a wrong classification as explained in Section~\ref{sec:sg}, we analyse only sources with \verb+NBANDS+ > 5. From the list of 2,761 extragalactic sources, 514 targets meet this requirement. In this sample, we identify two phenotypes of galaxies that mimic the LT colours: elliptical galaxies at redshifts $1. < z < 2.$ and another phenotype of spirals and irregulars with $5. < z < 6.$. In Fig.~\ref{fig:modvsphotoz} we show the best-fit galaxy template versus the redshift. In colour scale, the best-fit MLT type is presented. In Fig.~\ref{fig:nmod} we show the number of extragalactic sources as a function of galaxy spectral type. The contamination happens only in the ML regime.

\section{Validation}
\label{app:codeval}

As a simple validation test, the expected number counts output from \textit{GalmodBD} for a field with unit solid angle, in the case of a  uniform BD LF of $n_0 \ pc^{-3}$, and towards the Galactic poles may be compared to the trivial analytical solution for this particular case, which comes from direct integration of the fundamental equation of stellar statistics for a single disk exponential model:

\begin{multline}
N(m)= \\ n_0 [h_z~f_1(d_1,d_2) +2~h_z^2~f_2(d_1,d_2)+2~h_z^3~f_3(d_1,d_2)]
\label{eq:Nm}
\end{multline}


In the expression above, if $m,m+dm$ is some chosen bin in apparent magnitude, $N(m)dm$ will be the number of objects in the given bin. $h_z$ is the model exponential scale height for the disk, and 

\begin{equation}
f_1(d_1,d_2)=d_1^2~exp(-d_1/h_z) - d_2^2~exp(-d_2/h_z)
\end{equation}
\begin{equation}
f_2(d_1,d_2)=d_1~exp(-d_1/h_z) - d_2~exp(-d_2/h_z) 
\end{equation}
\begin{equation}
f_3(d_1,d_2)=exp(-d_1/h_z) - exp(-d_2/h_z) 
\end{equation}

where

\begin{equation}
d_1 = 10^{(0.2(m-M_{ft}+5))} \ pc 
\end{equation}
\begin{equation}
d_2 = 10^{(0.2(m-M_{br}+5))} \ pc 
\end{equation}

are, respectively, the minimum and maximum distances out to which any BD can be observed with the apparent magnitude $m$. This minimum (maximum) distance inevitably corresponds to the least (most) luminous BD spectral type in some filter, whose absolute magnitude is $M_{ft}$ ($M_{br}$).

Another simple validation test, again for the same special case as before, but this time involving all spectral types, is provided by the cumulative counts within some magnitude range $[m_{br},m_{ft}]$, $N(\leq m)$:

\begin{multline}
\label{eq:11}
N(\leq m) = \\
\sum_k^{Nmod} n_0 [h_z~f_1(k,d_1,d_2) +2~h_z^2~f_2(k,d_1,d_2)+ \\ 
2~h_z^3~f_3(k,d_1,d_2)]
\end{multline}

where in the expression above, the sum is over all $Nmod$ BD spectral types, and the functions $f_1$, $f_2$, and $f_3$ are as given before. The minimum and maximum distances are now given by

\begin{equation}
d_1 = 10^{(0.2(m_{br}-M_k+5))} \ pc 
\end{equation}
\begin{equation}
d_2= 10^{(0.2(m_{ft}-M_k+5))} \ pc 
\end{equation}

therefore corresponding to the minimum and maximum distances over which the k-th model contributes to the counts. 

Completely analogous analytic expressions apply to the situation in which we model the number counts for $(l,b)=(180, 0)\deg$. In this case, all one needs to do is to replace the model disk scale height $h_{z,thin}$ by the scale length $h_R$.

\begin{figure}
\includegraphics[width=\columnwidth]{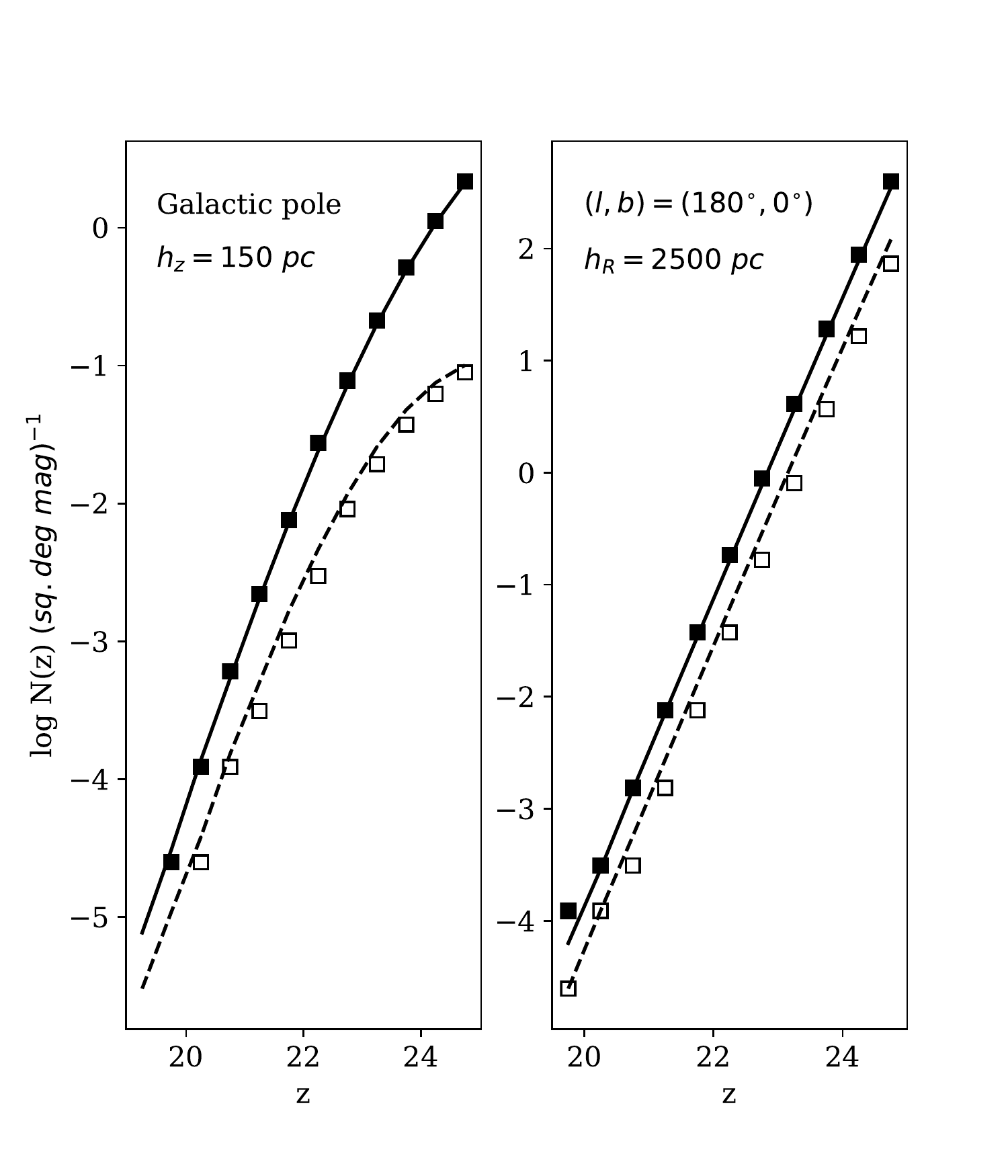}
\caption{Left panel: Star counts as a function of $z_{AB}$ towards the Galactic Pole using a single disk model with scale height $h_{z,thin} = 150 \ pc$. Points come from \textit{GalmodBD}: the open symbols are differential counts and the filled symbols are cumulative counts. The associated curves are the analytic formulae given by equations~\ref{eq:Nm} and ~\ref{eq:11}, respectively. Right panel: Same as in the previous panel, but now star counts towards the Galactic anti-centre, for a model with a single disk with horizontal scale length $h_R = 2.5 \ kpc$.}
  \label{fig:planaly}
\end{figure}

In Fig.~\ref{fig:planaly}, we show the comparison of \textit{GalmodBD} predictions with the analytical counts, based on a single disk component over a 1 sq. deg. field, and using $n_0$=$4~10^{-4} \ pc^{-3} mag^{-1}$, for the following cases: 1 - Galactic pole (left panel, using $h_z=150 \ pc$); 2-$(l,b)=(180,0,)\deg$ (right panel, with $h_R = 2500 \ pc$). The points are from \textit{GalmodBD} and the lines are from the analytic expressions provided above. Open (filled) symbols are differential (cumulative) counts as a function of $z$ band magnitude. \textit{GalmodBD} clearly reproduces the expected counts.

\bsp    
\label{lastpage}
\end{document}